\documentclass[12pt]{iopart}

\expandafter\let\csname equation*\endcsname\relax
\expandafter\let\csname endequation*\endcsname\relax

\usepackage{placeins}
\usepackage{amsfonts}
\usepackage{amssymb}
\usepackage{amsmath, mathtools}
\usepackage{graphicx} 
\usepackage{datetime}
\usepackage{color}
\usepackage{enumerate}
\usepackage{pifont,array}
\usepackage[dvipsnames]{xcolor}
\usepackage{cite}

\newcommand{\pdi}[2]{\frac{\partial#2}{\partial#1}}
\newcommand{\ket}[1]{|#1\rangle}
\newcommand{\bra}[1]{\langle#1|}
\newcommand{\expectn}[1]{\langle#1\rangle}

\begin{document}

\title{Near optimal discrimination of binary coherent signals via atom-light interaction}

\author{Rui Han$^{1,2}$\footnote{Present address: Centre for Quantum Technologies, National University of Singapore, 117543 Singapore}, J\'anos A. Bergou$^3$ and Gerd Leuchs$^{1,2,4}$}

\address{$^1$Max Planck Institute for the Science of Light, 91058 Erlangen, Germany}
\address{$^2$Institute of Optics, Information and Photonics, University of Erlangen-N\"urnberg, 91058 Erlangen, Germany}
\address{$^3$Department of Physics and Astronomy, Hunter College of the City, University of New York, 10065 New York, NY, USA} 
\address{$^4$University of Ottawa, Ottawa ON K1N 6N5, Canada}
\ead{han.rui@quantumlah.org}

\begin{abstract}

We study the discrimination of weak coherent states of light with significant overlaps by nondestructive measurements on the light states through measuring atomic states that are entangled to the coherent states via dipole coupling. In this way, the problem of measuring and discriminating coherent light states is shifted to finding the appropriate atom-light interaction and atomic measurements. We show that this scheme allows us to attain a probability of error extremely close to the Helstrom bound, the ultimate quantum limit for discriminating binary quantum states, through the simple Jaynes-Cummings interaction between the field and ancilla with optimized light-atom coupling and projective measurements on the atomic states. Moreover, since the measurement is nondestructive on the light state, information that is not detected by one measurement can be extracted from the post-measurement light states through subsequent measurements.

\end{abstract}

\noindent{\it Keywords\/}: coherent state, quantum communication, state discrimination, atom-light interaction, Helstrom bound, nondestructive measurement

\vspace{10ex}
\noindent{Journal Reference:} \emph{New J. Phys.} \textbf{20} 043005 (2018)
\maketitle

\section{Introduction}

Nonorthogonal coherent-state signal has become one of the most prominent quantum information carriers suitable for tasks such as quantum communication, sensing and cryptography.  Coherent signals have great advantage over others because they are easy to generate and have the best achievable signal-to-noise ratio during the information propagation. It is, however, challenging for the receiver to discriminate them in order to correctly decode the information, when the signals are weak and have significant overlaps. Most of the existing strategies of discriminating weak coherent states~\cite{Kennedy73,Dolinar1973,Bondurant:93,PhysRevLett.89.167901,Lorenz04,PhysRevLett.95.180503, PhysRevA.71.022318, PhysRevA.78.022320, Bergou:10, PhysRevLett.104.100505, PhysRevA.81.062338, RevModPhys.84.621,Becerra13, Becerra:14, PhysRevLett.117.200501, doi:10.1080/09500340903145031, PhysRevLett.106.250503, Muller2015}, including the displacement-controlled photon-number-resolving-detection (PNRD) strategies, rely on the receiver to perform a direct standard quantum projective measurement on the light state. 
Many schemes have demonstrated, with proof-of-principle experiments, that the standard quantum limit given by perfect Homodyne measurement can be surpassed.
The Dolinar-type receivers~\cite{Dolinar1973, Becerra:14} built upon the Kennedy receiver~\cite{Kennedy73} with real-time quantum feedback and highly nonlinear operations can achieve the Helstrom bound~\cite{Helstrom:76} in principle, but they are extremely difficult to implement in practice. 
The recently proposed strategy of Ref.~\cite{PhysRevLett.117.200501} replaces the feedback loop in Dolinar receiver by a feedforward loop and proves that the Helstrom bound can be asymptotically reached if the signal can be split into infinitely many individually accessible channels.
The implementations of these strategies are limited mainly by the quantum efficiency and dark count rate of the photon-number-resolving detectors (the highest demonstrated detection efficiency is about 91\%~\cite{PhysRevLett.106.250503, Calkins:13}), as well as the precision control of the optical-electrical loop for the real-time feedback.
As of yet, almost half a century after the proposal of the Dolinar receiver, there remains a significant gap between the practically achieved minimum error probability by (or even potentially achievable by) the existing schemes on discriminating coherent-state signals and the Helstrom bound -- the ultimate quantum limit.

In principle, when the Helstrom bound is not saturated, there could still be some information left in the system by generalized measurements (POVMs) or even in the case of projective measurements~\cite{PhysRevA.84.032326}. 
However, in practice, the light signal that enters a photon detector is completely destroyed regardless of the detection efficiency since there is no residue signal escaping from a conventional photon detector.
Therefore, information that is not accessed by such measurements is permanently lost which makes it impossible to reach the Helstrom bound. 

In this paper, in order to overcome the obstacles at the receiver's end, we explore the physical implementation to better discriminate binary coherent-state signals using the nondestructive measurement scheme proposed recently in Ref.~\cite{Han:2017a}. 
This scheme employs the Neumark dilation theorem for the implementation of a POVM~\cite{Neu40}. 
The key of this nondestructive implementation is to entangle the coherent light states with a two-level ancilla atom and discriminate the coherent states by measuring the state of the ancilla. This is equivalent to performing a two-element POVM measurement on the light signal.
The measurement is nondestructive since it is performed only on the ancilla such that the post-measurement light state is not destroyed.
In fact, this is a well-known method for many different quantum measurement problems~\cite{PhysRevLett.111.100501}, yet it has not been much explored for discriminating coherent states of light. 

The general scheme of Ref.~\cite{Han:2017a} is briefly described as follows. 
Alice prepares signal state $\{\ket{\psi_1}=\ket{\alpha},\ket{\psi_2}=\ket{\!-\!\alpha}\}$ with \emph{a priori} probabilities (referred to as priors for brevity) $\{\eta_1,\eta_2\}$ and sends it to Bob. Then, instead of performing measurements on the received state directly, Bob introduces an ancilla qubit initially prepared in state $\ket{\mathrm{i}}$ and entangles it with the signal state he received by some unitary transformation $U$. This procedure can be represented as
\begin{equation}
\begin{array}{l}U\ket{\psi_1}\ket{\mathrm{i}}=\sqrt{p_1}\ket{\varphi_1}\ket{1}+\sqrt{r_1}\ket{\phi_1}\ket{2}\,, \\
U\ket{\psi_2}\ket{\mathrm{i}}=\sqrt{r_2}\ket{\varphi_2}\ket{1}+\sqrt{p_2}\ket{\phi_2}\ket{2}\,,\end{array}\label{unitary}
\end{equation}
where $\{\ket{1},\ket{2}\}$ is an orthogonal basis of the ancilla qubit. The nondestructive measurement on the state Bob received is the projective measurements $\{\ket{1}\bra{1},\ket{2}\bra{2}\}$ on the ancilla qubit. 
If the unitary transformation relates state $\ket{\varphi_1}$ to state $\ket{1}$ and state $\ket{\phi_2}$ to state $\ket{2}$ with amplitudes as large as possible, Bob identifies the measurement result of $\ket{1}\bra{1}$ with state $\ket{\psi_1}$ and the measurement result of $\ket{2}\bra{2}$ with state $\ket{\psi_2}$. 
Hence, the error probability is
\begin{equation}
P_\mathrm{err}= \eta_1r_1+ \eta_2r_2\,.\label{eq:PerrJanos}
\end{equation}
It is shown that in the case where the post-measurement states no longer carry any information, i.e., $\ket{\varphi_1}=\ket{\varphi_2}$ and $\ket{\phi_1}=\ket{\phi_2}$, the Helstrom bound is reached when
\begin{equation}
r_{1,2}=\frac{1}{2}\left(1-\frac{1-2\eta_{2,1}s^2}{\sqrt{1-4 \eta_1 \eta_2s^2}}\right),
\end{equation}
where $s=|\expectn{\psi_1|\psi_2}|$ is the overlap between the two signal states.

On the other hand, the error probability for Bob's guess is higher than the Helstrom bound if the post-measurement states are different, i.e., $\ket{\varphi_1}\neq\ket{\varphi_2}$ and/or $\ket{\phi_1}\neq\ket{\phi_2}$. In this case, Bob can perform subsequent measurements on the post-measurement states to attain more information about the state sent by Alice. 
The subsequent measurements can be either projective or nondestructive. With a chain of nondestructive subsequent measurements, the scheme can be illustrated as
\begin{equation}
\left\{\begin{array}{c} \ket{\psi_1} \vspace{1ex}\\ \ket{\psi_2}\end{array}\right.\xrightarrow[\scriptsize{\color{red}\begin{array}{c}\mathrm{1st}\\\mathrm{measurement}\end{array}}]{U,\;\ket{\mathrm{i}}} 
\left\{\begin{array}{c} \ket{\psi_1^{(1)}} \vspace{1ex}\\ \ket{\psi_2^{(1)}}\end{array}\right.\xrightarrow[\scriptsize{\color{red}\begin{array}{c}\mathrm{2nd}\\\mathrm{measurement}\end{array}}]{U^{(1)},\;\ket{\mathrm{i}^{(1)}}}
\left\{\begin{array}{c} \ket{\psi_1^{(2)}} \vspace{1ex}\\ \ket{\psi_2^{(2)}}\end{array}\right.\xrightarrow[\mathrm{\color{red}\dots}]{}\cdots,
\end{equation}
where the post-measurement states $\{\ket{\psi_1^{(k)}},\ket{\psi_2^{(k)}}\}$ depend on the choice of the unitary operations $U^{(k-1)}$, the ancilla states and the previous measurement outcomes. This scheme becomes extremely useful when the requirement of reaching the Helstrom bound by a single measurement, such as producing identical post-measurement states, cannot be fulfilled in a realistic physical system. The information gain from each nondestructive measurement can be optimized by choosing a suitable unitary transformation $U^{(k)}$ and atomic measurements. In the ideal scenario of fast convergence of the error probability, only a few rounds of the nondestructive measurements would be sufficient. 

This paper is organized as follows. We first describe, in Section 2, the physical system employed in our implementation scheme, i.e., the light-atom system with the Jaynes-Cummings interaction. 
In Section 3, we illustrate the procedure of finding the optimal measurements and optimal coupling for both cases of equal and biased prior distributions. We show that the error probability of our scheme can be extremely close to the Helstrom bound. 
The sequential measurement scheme is also briefly discussed at the end of the section.
In Section 4, we discuss first the fundamental difference between this present scheme and the existing ones that leads to its advance in reaching the Helstrom bound and then its possible extensions. We close with a short summary in Section 5.

\section{The Physical System}

The system of a single-mode coherent light field interacting with a two-level atom is described by the Jaynes-Cummings model~\cite{1443594} with its Hamiltonian given by
\begin{equation}
H=\hbar \omega_L a^\dagger a+\frac{1}{2}\hbar\omega_0\sigma_z+\hbar g(\sigma_+a+a^\dagger\sigma_-)\,,
\end{equation}
where $\omega_L$ is the frequency of the light field, $\omega_0$ is the frequency of the atomic transition, $\sigma_+$ and $\sigma_-$ denote the atomic raising and lowering, $a$ and $a^\dag$ are the field annihilation and creation operators. The vacuum coupling strength $g$ depends on the properties and confinement of the light field and the dipole moment of the atomic transition. 
When the dipole coupling is on resonance, i.e., $\omega_L=\omega_0$, the Hamiltonian in the interaction picture is simply reduced to
\begin{equation}
H_I=\hbar g(\sigma_+a+\sigma_- a^\dagger).\label{eq:AtomLightInteraction}
\end{equation}
The total Hilbert space is a tensor product space of the two-level atom and the light field spanned by $\{\ket{\mathrm{g},n},\ket{\mathrm{e},n}, \;\mathrm{for}\;n=0,1,2,\cdots\}$, where $\ket{n}$ denotes the Fock state with $n$ photons. 

Here, the ancilla qubit states $\ket{\mathrm{g}}$ and $\ket{\mathrm{e}}$ can be the ground and excited states of a two-level atom that is resonantly coupled to an optical field with frequency $\omega_L$. In this case, the interaction Hamiltonian of Eq.~(\ref{eq:AtomLightInteraction}) is a good description of the system when the dissipation from the excited state is much weaker than the coupling strength $g$. On the other hand, the dissipation effect can be neglected in an effective two-level description of the three-level Raman transition of the $\Lambda$-configuration where the two `ground' states labeled by $\ket{\mathrm{g}}$ and $\ket{\mathrm{e}}$ are connected by a two-photon transition via a far-detuned intermediate state. Eq.~(\ref{eq:AtomLightInteraction}) would be a good effective interaction Hamiltonian between the two `ground' states $\ket{\mathrm{g}}$ and $\ket{\mathrm{e}}$ if $\ket{\mathrm{g}}$ (or $\ket{\mathrm{e}}$) is coupled to the intermediate state with a strong pumping field and $\ket{\mathrm{e}}$ (or $\ket{\mathrm{g}}$) is coupled to the intermediate state with the weak optical signal of our interest. The effective coupling strength $g$ between $\ket{\mathrm{g}}$ and $\ket{\mathrm{e}}$ can then also be controlled by adjusting the pumping field.
Moreover, in order to increase the coupling efficiency, one can also employ an atomic ensemble (such as a Bose-Einstein condensate) or an artificial atom (such as a quantum dot) as the ancilla qubit. 
The presented scheme works as long as the interaction between the ancilla and the field can be described by Eq.~\eref{eq:AtomLightInteraction}.
Although the free-space coupling between atoms and light is typically very weak, cavity quantum electrodynamics shows that this coupling can be enhanced by orders of magnitude when the interaction is confined in a cavity. For examples, the vacuum coupling strength $g$ can be up to a few hundreds of MHz for the interaction between a trapped rubidium atom and optical light field~\cite{nature09997}; and, $g=24\times2\pi$~GHz can be reach for the coupling between light and collective states of a Bose-Einstein condensate~\cite{nature06331}.

The evolution of state is given by the solution of the pairwise coupled Schr\"odinger's equations of motion in the interaction picture,
\begin{equation}
i\hbar\pdi{t}{}\left(\begin{array}{c}c_{\mathrm{g},n}(t)\\c_{\mathrm{e},n-1}(t)\end{array}\right)=\hbar\sqrt{n}g(t)\left(\begin{array}{cc}0&1\\1&0\end{array}\right)\left(\begin{array}{c}c_{\mathrm{g},n}(t)\\c_{\mathrm{e},n-1}(t)\end{array}\right),
\end{equation}
where $c_{\mathrm{g},n}(t)$ and $c_{\mathrm{e},n}(t)$ denote the coefficients of the atom-light state at a later time $t$,
\begin{equation}
\ket{\Psi(t)}=\sum_{n=0}^\infty\Big(c_{\mathrm{g},n}(t)\ket{\mathrm{g},n}+c_{\mathrm{e},n}(t)\ket{\mathrm{e},n}\Big).
\end{equation}
Solutions to these equations are
\begin{equation}
c_{\mathrm{g},n}(t)\pm c_{\mathrm{e},n-1}(t)=e^{\mp i\sqrt{n}\Phi(t)}[c_{\mathrm{g},n}(0)\pm c_{\mathrm{e},n-1}(0)]\,,\label{eq:cgcent}
\end{equation}
which depend on the time integrated coupling strength
\begin{equation}
\Phi(t)=\int_0^t dt' \,g(t').\label{eq:Phi}
\end{equation}
Since coefficients of the state at time $t$ given by Eq.~(\ref{eq:cgcent}) solely depend on $\Phi(t)$, we can regard the state as a function depending on a single free parameter $\Phi(t)$. Therefore, for brevity, we can also omit the time dependence in $\Phi(t)$ and just denote it by $\Phi$ and represent
\begin{equation}
c_{\mathrm{g},n}(t)\rightarrow c_{\mathrm{g},n}(\Phi)\;\;\;\mathrm{and}\;\;\;c_{\mathrm{e},n}(t)\rightarrow c_{\mathrm{e},n}(\Phi)\,.\nonumber
\end{equation}
However, one should not forget that $\Phi$ explicitly depends on time $t$, the dipole-dipole coupling strength, the temporal profile of the field, etc. 

If the initial light field is in a coherent state given by
\begin{equation}
\ket{\alpha}=\sum_{n=0}^\infty\alpha_n\ket{n}\;\;\;\mathrm{with}\;\;\;\alpha_n=e^{-|\alpha|^2/2}\frac{\alpha^n}{\sqrt{n!}}\,
\end{equation}
and interacting with an atom in its ground state $\ket{\mathrm{i}}=\ket{\mathrm{g}}$, i.e., $c_{\mathrm{g},n}(0)=\alpha_n$ and $c_{\mathrm{e},n}(0)=0$,
the state of the system at a later time $t$ is~\cite{PhysRevLett.44.1323, PhysRevLett.58.353, PhysRevA.44.6023, Shore:93}
\begin{equation}
\ket{\Psi(t)}=\sum_{n=0}^\infty\Big[\hspace{-2.5pt}\cos(\Phi\sqrt{n})\alpha_n\ket{\mathrm{g},n}-i\sin(\Phi\sqrt{n+1})\alpha_{n+1}\ket{\mathrm{e},n}\Big].\label{eq:Psit1}
\end{equation}
In general, Bob's decision on the light signal depends on the outcome of the measurement on the atomic state $\rho_A$, which is obtained by tracing out the light field
\begin{equation}
\rho_A(t)=\mathrm{tr}_L\{\ket{\Psi(t)}\bra{\Psi(t)}\}\,,
\end{equation}
and the post-measurement light state depends on both the measurement operators and the respective outcome. In this case, the atomic state is
\begin{equation}
\rho_{A,\alpha}(t)=\sum_{n=0}^\infty\left(\begin{array}{cc}\cos^2(\Phi\sqrt{n})|\alpha_n|^2 & i\cos(\Phi\sqrt{n})\sin(\Phi\sqrt{n+1})\alpha_n\alpha^*_{n+1} \\ -i\cos(\Phi\sqrt{n})\sin(\Phi\sqrt{n+1})\alpha^*_n\alpha_{n+1} & \sin^2(\Phi\sqrt{n+1})|\alpha_{n+1}|^2\end{array}\right).\label{eq:AstateAlpha}
\end{equation}

\section{The minimum error state discrimination}

The maximum distinguishability between two quantum states $\rho_1$ and $\rho_2$ is related to the trace distance:
\begin{equation}
D_\mathrm{tr}(\rho_1,\rho_2)\equiv \frac{1}{2}||\rho_1-\rho_2||_\mathrm{tr}=\frac{1}{2}\sum_j|\lambda_j|,
\end{equation}
where $\lambda_j$ is the $j$th eigenvalue of the Hermitian operator $\rho_1-\rho_2$. Connecting to quantum measurements, the trace distance between any two density operators multiplied with their priors can be expressed as
\begin{equation}
D_\mathrm{tr}(\eta_1\rho_1,\eta_2\rho_2)= \underset{\Pi}{\mathrm{max}}\big|\mathrm{tr}\{{\Pi(\eta_1\rho_1-\eta_2\rho_2)}\}\big|-\frac{1}{2}|\eta_1-\eta_2|,\label{eq:trDPi2}
\end{equation}
where the maximization is over all possible projective measurements $\Pi$. In the nondestructive measurement scheme, Bob makes a decision upon the measurement outcomes of the ancilla states $\{\rho_{A,\alpha},\rho_{A,-\alpha}\}$ with prior probabilities $\{\eta_1,\eta_2\}$. Thus, Bob's error probability for discriminating coherent signals $\{\ket{\alpha},\ket{{-}\alpha}\}$ is bounded by
\begin{equation}
P_\mathrm{err}^{\mathrm{min}}=\min_{\Phi}\frac{1}{2}\left[1-2D_\mathrm{tr}(\eta_1\rho_{A,\alpha},\eta_2\rho_{A,-\alpha})\right],\label{eq:PerrMin1}
\end{equation}
where the minimization is over the time integrated atom-light interaction strength $\Phi(t)$ that can be adjusted experimentally. 

\begin{figure}[t]
\centerline{
\begin{picture}(230,130)(0,0)
\put(94,5){\includegraphics[scale=0.5]{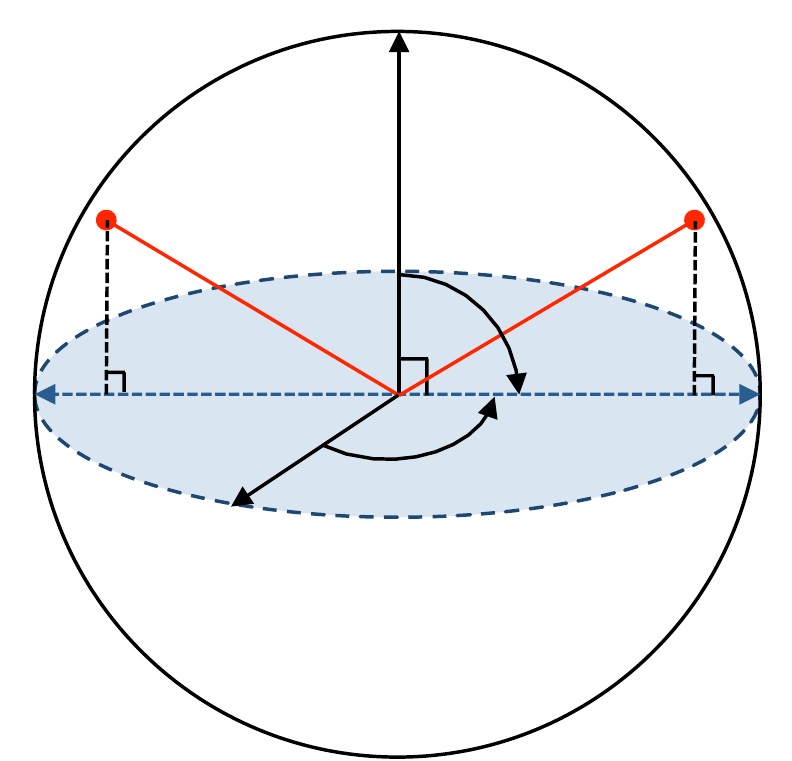}}
\put(-12,28){\includegraphics[scale=0.46]{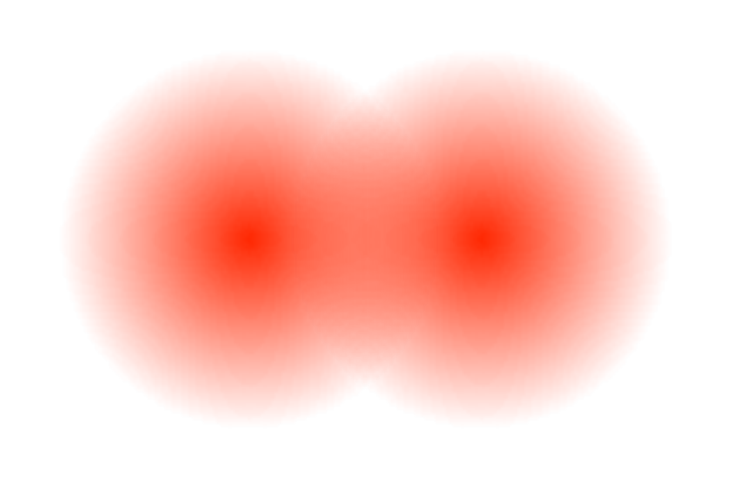}}
\put(220,58){\makebox(0,0){$\hat{\boldsymbol{y}}$, $\Pi_+$}}
\put(90,58){\makebox(0,0){$\Pi_-$}}
\put(154,120){\makebox(0,0){$\hat{\boldsymbol{z}}$, $\ket{\mathrm{g}}$}}
\put(154,-3){\makebox(0,0){$\ket{\mathrm{e}}$}}
\put(124,38){\makebox(0,0){$\hat{\boldsymbol{x}}$}}
\put(95,87){\makebox(0,0){$\rho_{A,\alpha}$}}
\put(210,87){\makebox(0,0){$\rho_{A,-\alpha}$}}
\put(56,60){\makebox(0,0){$\ket{\alpha}$}}
\put(20,60){\makebox(0,0){$\ket{\!-\!\alpha}$}}
\put(160,46){\makebox(0,0){$\theta$}}
\put(173,70){\makebox(0,0){\footnotesize{$2\arctan\gamma$}}}
\end{picture}}
\caption{Scheme of the nondestructive implementation for the discrimination of binary coherent-state signals $\{\ket{\alpha},\ket{\!-\!\alpha}\}$ with equal priors. The ancilla atom is initially prepared in $\ket{\mathrm{g}}$ with its Bloch vector pointing along the positive $\hat{\boldsymbol{z}}$ direction. The Jaynes-Cummings interaction with coherent state $\ket{\pm\alpha}$ rotates the atomic Bloch vector symmetrically about $\hat{\boldsymbol{x}}$ to mixed state $\rho_{A,\pm\alpha}$.}\label{fig1}
\end{figure}
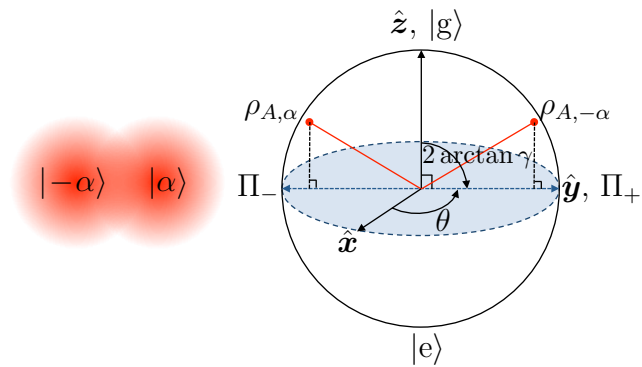

In this section, we evaluate Bob's minimum error probability with a fixed initial ancilla state $\ket{\mathrm{i}}=\ket{\mathrm{g}}$ where the atomic states $\rho_{A,\pm\alpha}$ are given by Eq.~(\ref{eq:AstateAlpha}). Since $\alpha_n\alpha^*_{n+1}=|\alpha_n|^2\alpha^*/\sqrt{n+1}$, the two density operators $\rho_{A,\alpha}(t)$ and $\rho_{A,-\alpha}(t)$ differ only by an overall sign of their off-diagonal elements. In this case, the Jaynes-Cummings interaction with coherent states $\ket{\pm\alpha}$ gives rise to a $\sigma_x$ rotation of the atomic state $\ket{\mathrm{g}}$ to $\rho_{A,\pm\alpha}$ (see Fig.~\ref{fig1}). $\rho_{A,\pm\alpha}$ are mixed states as a result of the entanglement to the light field, and the rotation of the atomic state is symmetric with respect to the states $\ket{\alpha}$ and $\ket{\!-\!\alpha}$.

The projective measurement onto an arbitrary pure state $(\ket{\mathrm{g}}+e^{i\theta}\gamma\ket{\mathrm{e}})/\sqrt{1+\gamma^2}$ with real parameters $\gamma$ and $\theta$ can be generally represented as
\begin{equation}
\Pi(\gamma,\theta)=\frac{1}{1+\gamma^2}\left(\begin{array}{cc}1&\gamma e^{-i\theta}\\\gamma e^{i\theta}&\gamma^2\end{array}\right)
\end{equation}
in the basis of $\{\ket{\mathrm{g}},\ket{\mathrm{e}}\}$. 
In order to reach $P_\mathrm{err}^{\mathrm{min}}$ in (\ref{eq:PerrMin1}), we need to find measurement $\Pi(\gamma,\theta)$ that optimally discriminate states $\rho_{A,\pm\alpha}$ as well as the optimal atom-light interaction parameter $\Phi$ that gives the maximum distinguishability of $\rho_{A,\pm\alpha}$.

\subsection{Equal priors $\eta_1=\eta_2$}

For signals with equal priors, Bob's error probability with projective measurement $\Pi(\gamma,\theta)$ on the ancilla atom is $P_\mathrm{err}=\frac{1}{2}\big[1-\big|\mathrm{tr}\{{\Pi(\gamma,\theta)(\rho_{A,\alpha}-\rho_{A,-\alpha})}\}\big|\big]$, which strongly depends on the atom-light interaction parameter $\Phi$ and the choice of atomic measurement $\Pi(\gamma,\theta)$. To minimize $P_\mathrm{err}$, we will first seek for the optimal measurement operator which extracts maximum knowledge of the atomic state, i.e., $\underset{\Pi}{\mathrm{max}}\big|\mathrm{tr}\{{\Pi(\gamma,\theta)(\rho_{A,\alpha}-\rho_{A,-\alpha})}\}\big|=D_\mathrm{tr}(\rho_{A,\alpha},\rho_{A,-\alpha})$, and then find the optimal interaction that maximizes the trace distance of the atomic states $D_\mathrm{tr}(\rho_{A,\alpha},\rho_{A,-\alpha})$ which depends on a single parameter $\Phi$ for the atomic state given in Eq.~(\ref{eq:AstateAlpha}).

\subsubsection{The optimal measurement}

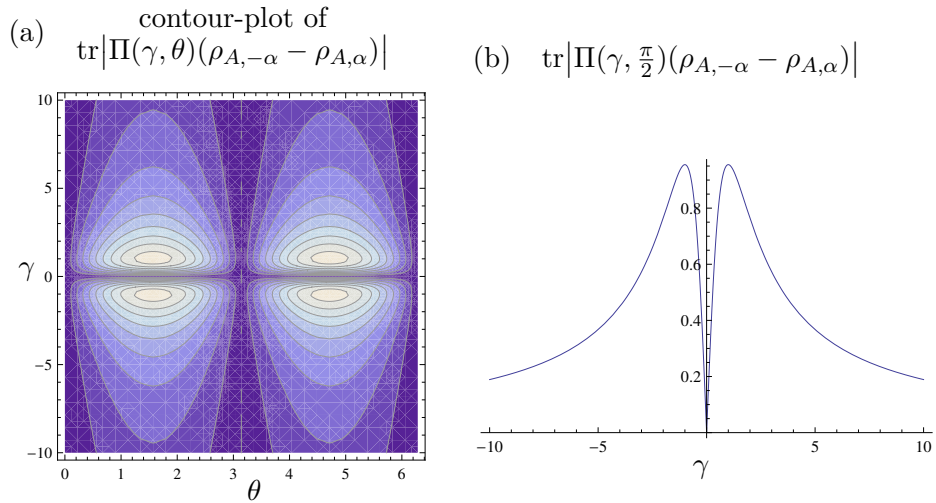
\begin{figure}[t]
\footnotesize
\centerline{\setlength{\unitlength}{1pt}
\begin{picture}(260,200)(0,0)
\put(0,0){\includegraphics[scale=0.65]{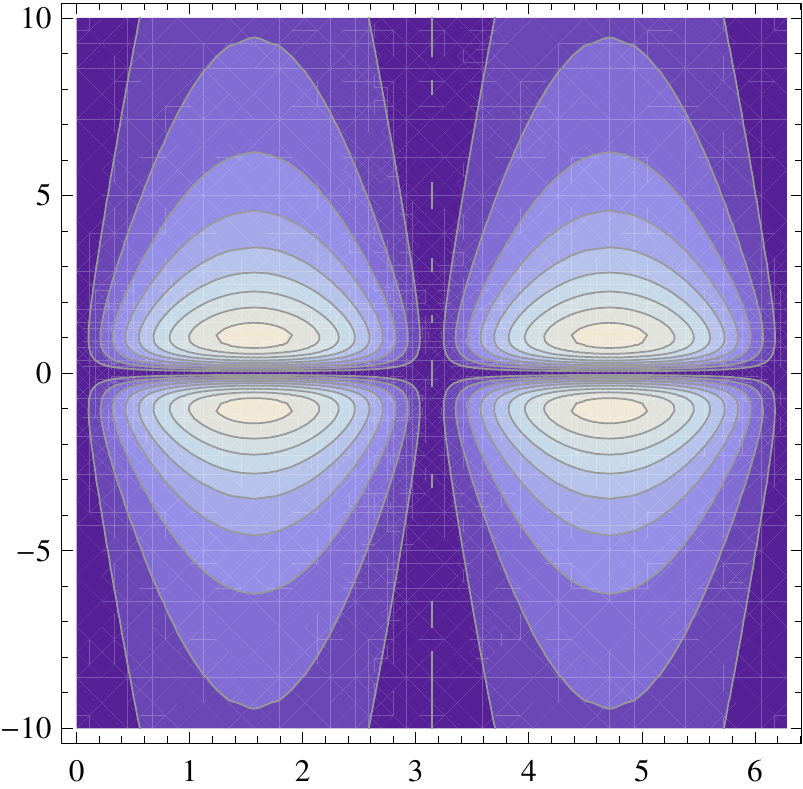}}
\put(170,10){\includegraphics[scale=0.68]{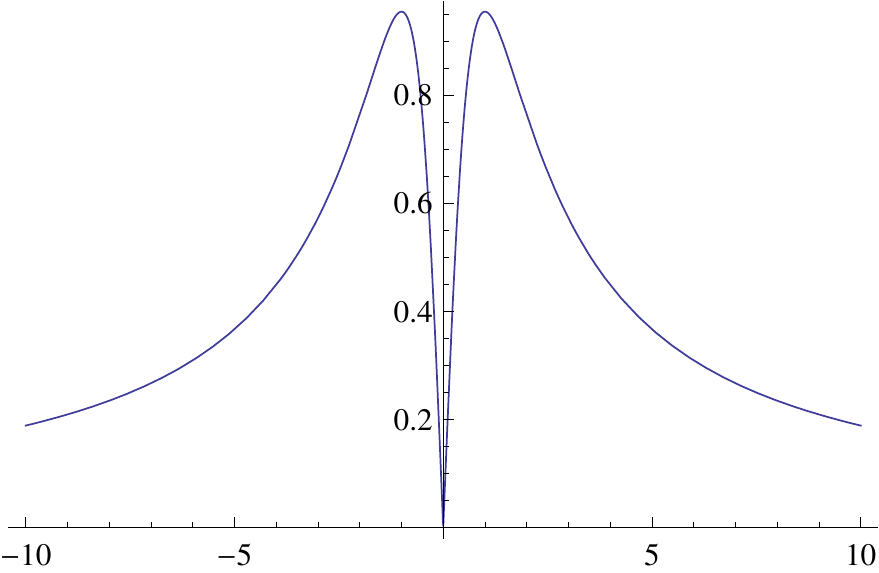}}
\put(0,170){\makebox(0,0){\small{(a)}}}
\put(176,160){\makebox(0,0){\small{(b)}}}
\put(78,175){\makebox(0,0){\small{contour-plot of}}}
\put(78,163){\makebox(0,0){\small{$\mathrm{tr}\big|\Pi(\gamma,\theta)(\rho_{A,-\alpha}-\rho_{A,\alpha})\big|$}}}
\put(255,160){\makebox(0,0){\small{$\mathrm{tr}\big|\Pi(\gamma,\frac{\pi}{2})(\rho_{A,-\alpha}-\rho_{A,\alpha})\big|$}}}
\put(255,5){\makebox(0,0){\small{$\gamma$}}}
\put(86,-3){\makebox(0,0){\small{$\theta$}}}
\put(0, 80){\makebox(0,0){\small{$\gamma$}}}
\end{picture}}
\caption{For weak signals with $\alpha=1$, (a) the contour-plot of the trace distance $\mathrm{tr}\big|{\Pi(\gamma,\theta)(\rho_{A,-\alpha}-\rho_{A,\alpha}})\big|$ vs. $\gamma$ and $\theta$, and (b) the trace distance vs. $\gamma$ when $\theta=\pi/2$. The numerically calculated values of $\Phi$ that maximize the RHS of~(\ref{eq:TrDpm}) are used for plotting the function. The figure indicates that $\mathrm{tr}\big|{\Pi(\gamma,\theta)(\rho_{A,-\alpha}-\rho_{A,\alpha}})\big|$ is maximal at $e^{i\theta}=\pm i$ and $\gamma=\pm1$.}\label{fig:TrDContourBeta}
\end{figure}

The knowledge of Bob obtained through an arbitrary projective measurement $\Pi(\gamma,\theta)$ on the ancilla is
\begin{equation}
\mathrm{tr}\big|\Pi(\gamma,\theta)(\rho_{A,\alpha}-\rho_{A,-\alpha})\big|=\underbrace{\frac{4\gamma\mathrm{Im}(e^{i\theta}\alpha)}{1+\gamma^2}}_{\leq2|\alpha|}\sum_{n=0}^\infty\frac{|\alpha_n|^2}{\sqrt{n+1}}\cos(\Phi\sqrt{n})\sin(\Phi\sqrt{n+1})\,.
\end{equation}
This knowledge strongly depends on parameters $\gamma$ and $\theta$; see Fig.~\ref{fig:TrDContourBeta}. Its upper bound is reached when $\gamma=1$ and Im$(e^{i\theta}\alpha)=|\alpha|$.
We note that the maximum value does not depend on the argument of $\alpha$ but only on its absolute value. Therefore, without losing its generality, we will consider only real values of $\alpha$ throughout the rest of this paper, because any complex phase of $\alpha$ can be compensated by changing angle $\theta$ in the atomic measurement. 

Hence, for real $\alpha$, the optimal measurements are $\Pi(1,\pm\pi/2)$, i.e.,
\begin{equation}
\Pi_+=\ket{+}\bra{+}\;\;\mathrm{and}\;\;\Pi_-=\ket{-}\bra{-},
\end{equation}
where $\ket{\pm}=\frac{1}{\sqrt{2}}(\ket{\mathrm{g}}\pm i\ket{\mathrm{e}})$. Moreover, it is very plausible that the optimal measurement does not depend on $\Phi$.
The probability of finding measurement outcome $\Pi_\pm$ on the states is
\begin{equation}
\tr{\rho_{A,\alpha}\Pi_\pm}=\frac{1}{2}\mp\alpha\sum_{n=0}^\infty\frac{\alpha_n^2}{\sqrt{n+1}}\cos(\Phi\sqrt{n})\sin(\Phi\sqrt{n+1}).
\end{equation}
Bob's strategy is to simply associate the detector click of $\Pi_+$ with light state $\ket{\!-\!\alpha}$ and the detector click of $\Pi_-$ with light state $\ket{\alpha}$. One can also simply verify that the trace distance between states $\rho_{A,\pm\alpha}$,
\begin{equation}
D_\mathrm{tr}(\rho_{A,\alpha},\rho_{A,-\alpha})=\left|2\alpha\sum_{n=0}^\infty\frac{\alpha_n^2}{\sqrt{n+1}}\hspace{-1pt}\cos(\Phi\sqrt{n})\sin(\Phi\sqrt{n+1})\right|,\label{eq:TrDpm}
\end{equation}
equals to the difference between the measurement results of $\Pi_\pm$.

\subsubsection{The minimum error probability}

The trace distance between the two atomic states $D_\mathrm{tr}(\rho_{A,\alpha},\rho_{A,-\alpha})$, given explicitly in Eq.~(\ref{eq:TrDpm}), is an oscillatory function of its dimensionless variable $\Phi$. 
Examples illustrating the time-dependence of this trace distance are given in Fig.~3 for $\alpha=\{2,1,0.5\}$.
Bob's error probability can be minimized by finding the optimal value of $\Phi$ that gives the maximum distinguishability.

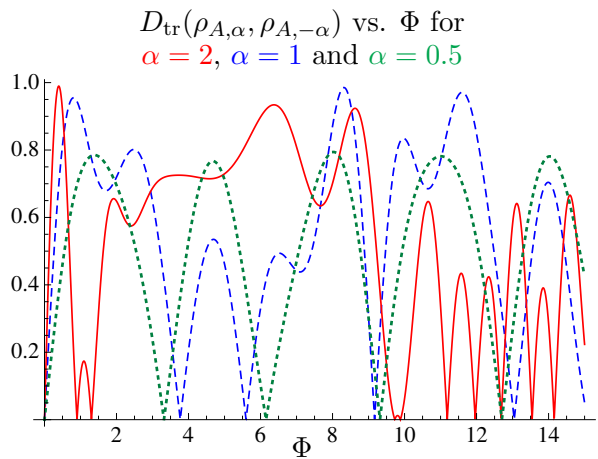
\begin{figure}[t]
\footnotesize
\centerline{\setlength{\unitlength}{1pt}
\begin{picture}(250,170)(0,0)
\put(5,0){\includegraphics[scale=0.65]{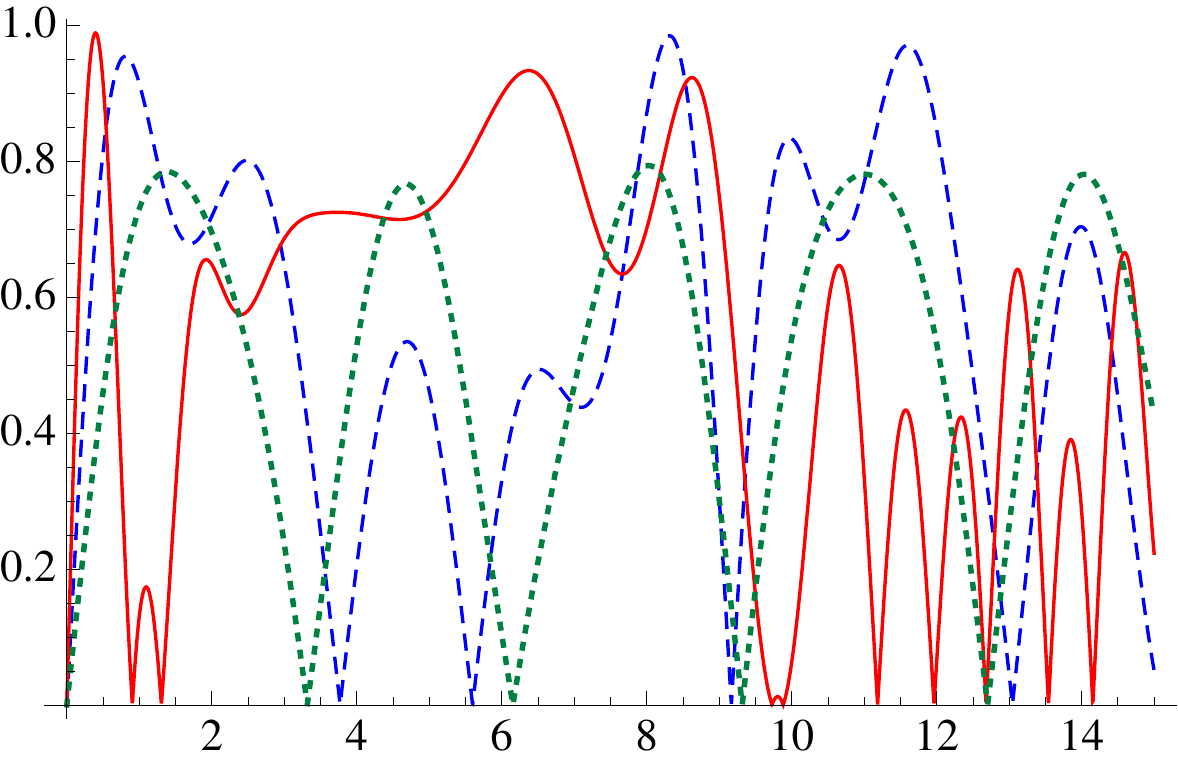}}
\put(115,160){\makebox(0,0){\small{$D_\mathrm{tr}(\rho_{A,\alpha},\rho_{A,-\alpha})$ vs. $\Phi$ for}}}
\put(115,148){\makebox(0,0){\small{\textcolor{red}{$\alpha=2$}, \textcolor{blue}{$\alpha=1$} and \textcolor{Green}{$\alpha=0.5$}}}}
\put(115,0){\makebox(0,0){\small{$\Phi$}}}
\end{picture}}
\caption{Trace distance $D_\mathrm{tr}(\rho_{A,\alpha},\rho_{A,-\alpha})$ against $\Phi$. The solid, dashed and dotted curves represent the trace distances for $\alpha=2$, $\alpha=1$ and $\alpha=0.5$, respectively.}\label{fig:TrDPhi}
\end{figure}

Although there are infinitely many local maxima of the function as $\Phi$ goes from 0 to $\infty$, the global maximum is either given by the first maximum for the smallest value of $\Phi$ or the maximum around $\Phi=8$. When $\alpha$ is large, the first local maximum of the function $D_\mathrm{tr}(\rho_{A,\alpha},\rho_{A,-\alpha})$ in $\Phi$ can be very close to unity, and it is also the global maximum; for example, when $\alpha=2$, the global maximum $D_\mathrm{tr}(\rho_{A,\alpha},\rho_{A,-\alpha})=0.9896$ is obtained for $\Phi\rightarrow0.3960$. As $\alpha$ gets smaller, the local maximum of the function around $\Phi=8$ becomes slightly larger than its first maximum; for example, when $\alpha=1$, these two maximum values are $0.9550$ for $\Phi\rightarrow0.8069$ and $0.9853$ for $\Phi\rightarrow8.3168$. As $\alpha$ gets smaller and smaller, the values of these local maxima get closer and closer; for example, when $\alpha=0.5$, $D_\mathrm{tr}(\rho_{A,\alpha},\rho_{A,-\alpha})=0.7941$ for $\Phi\rightarrow1.3857$ and $D_\mathrm{tr}(\rho_{A,\alpha},\rho_{A,-\alpha})=0.7851$ for $\Phi\rightarrow8.0285$. 
For every fixed value of $\alpha$, the optimal value of $\Phi$ can be evaluated numerically. 
Experimentally, one can reach the optimal value of $\Phi$ by controlling the duration of the light pulse and the atom-light coupling strength. 

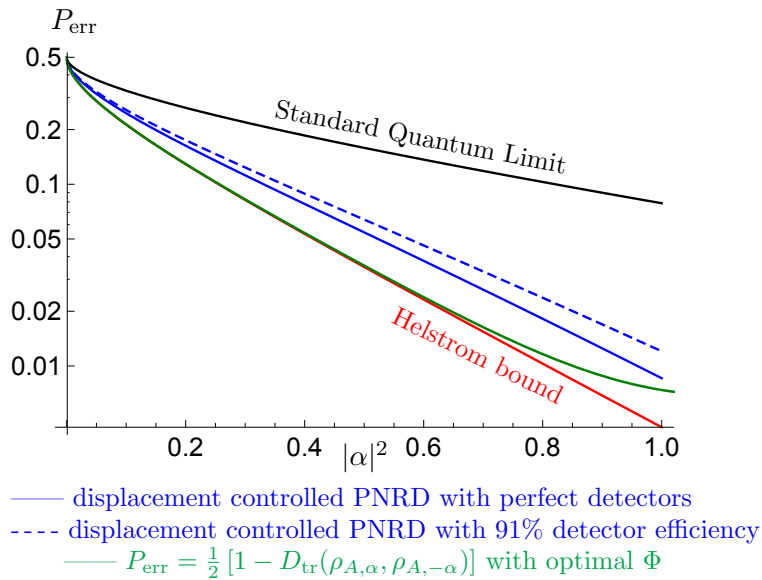
\begin{figure}[t]
\footnotesize
\centerline{\setlength{\unitlength}{1pt}
\begin{picture}(250,220)(0,0)
\put(0,40){
\put(0,0){\includegraphics[scale=0.7]{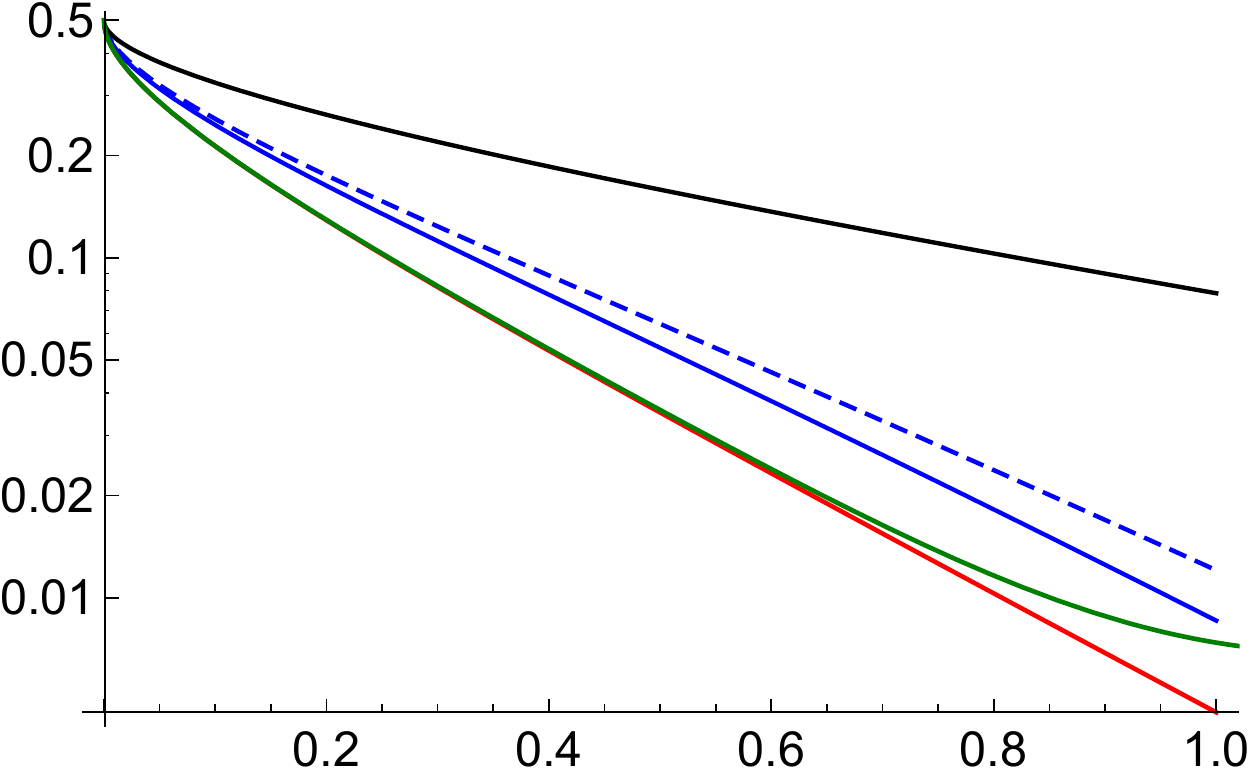}}
\put(125,0){\makebox(0,0)[l]{\small{$|\alpha|^2$}}}
\put(15,165){\makebox(0,0)[l]{\small{$P_\mathrm{err}$}}}
\put(100,131){\rotatebox{-11}{\makebox(0,0)[l]{Standard Quantum Limit}}}
\put(145,54){\rotatebox{-26}{\makebox(0,0)[l]{\color{red}Helstrom bound}}}
}
\put(0,24){\makebox(0,0)[l]{\color{blue}------ displacement controlled PNRD with perfect detectors}}
\put(0,12){\makebox(0,0)[l]{\color{blue}-\,-\,-\,- displacement controlled PNRD with 91\% detector efficiency}}
\put(20,0){\makebox(0,0)[l]{\color{Green}------ $P_\mathrm{err}=\frac{1}{2}\left[1-D_\mathrm{tr}(\rho_{A,\alpha},\rho_{A,-\alpha})\right]$ with optimal $\Phi$}}
\end{picture}}
\caption{Error probability vs. $|\alpha|^2$ for the discrimination of coherent states $\{\ket{\alpha},\ket{\!-\!\alpha}\}$ with equal prior probabilities. The vertical axis is shown in logarithm scale. Green curve: minimum error probability for the nondestructive measurement scheme with an ancilla atom initially prepared in $\ket{\mathrm{g}}$ and measured in the $\{\ket{+},\ket{-}\}$ basis after the optimized interaction with the light field; it is almost indistinguishable from the Helstrom bound shown by the red curve for small $|\alpha|^2$. }\label{fig:ErrorProbMin}
\end{figure}

Bob's minimum error probability under perfect experimental control is shown in Fig.~\ref{fig:ErrorProbMin}. 
It is evident from the figure that the minimum error probability of the present scheme gets extremely close to the Helstrom bound for discriminating coherent states with significant overlaps, i.e., for small values of $\alpha$. 
The discrepancy from the Helstrom bound is less than 0.1\% for $\alpha<0.85$ and less than 0.01\% for $\alpha<0.3$.
This range of coherent states with few photons is exactly the range of interest for a secure quantum communication channel.
For these states of applicational significance, our scheme by-far outperforms the perfect displacement-controlled PNRD strategy.
For the discrimination of weak binary coherent signals, this is the first physically implementable scheme that can come so close to the Helstrom bound without real-time quantum feedback.
Similar as the Dolinar-type receivers for PNRD strategies, quantum feedback loops can also further reduce the error probability of this scheme by splitting the signal and discriminating the weaker signals with adaptive measurements. Such feedback loops can be helpful for the discrimination of stronger signals where the discrepancy to the Helstrom bound is larger and/or the atom-light coupling efficiency is low.

Here, it is worth to again emphasize that the implementation of our scheme requires only the Jaynes-Cummings interaction with a two-level atom and an atomic measurement projected onto states $\ket{\pm}$. Both the Jaynes-Cummings Hamiltonian and the atomic projective measurement are experimentally implementable with very high precision~\cite{RevModPhys.73.565, Chu2002, Reiserer2015}. The only parameter that requires particularly careful control is the time integrated interaction strength $\Phi$.

\subsection{Unequal priors $ \eta_1\neq \eta_2$}

The relation between the trace distance and the measurement $\Pi$ for any two quantum states with an arbitrary prior distribution is given in Eq.~(\ref{eq:trDPi2}).
Although the symmetry of the problem is partially broken when the prior distribution is biased, the procedure used to tackle the problem of equal priors still applies. Thus, we will first seek for the optimal atomic measurement $\Pi(\gamma,\theta)$ that maximumly discriminates the atomic states, and then, find $P_\mathrm{err}^\mathrm{min}$ of Eq.~(\ref{eq:PerrMin1}) by maximizing the trace distance $D_\mathrm{tr}(\eta_1\rho_{A,\alpha},\eta_2\rho_{A,-\alpha})$ over $\Phi$.

\begin{figure}[t]
\footnotesize
\centerline{\setlength{\unitlength}{1pt}
\begin{picture}(240,300)(0,0)
\put(0,136){
\put(0,0){\includegraphics[scale=0.58]{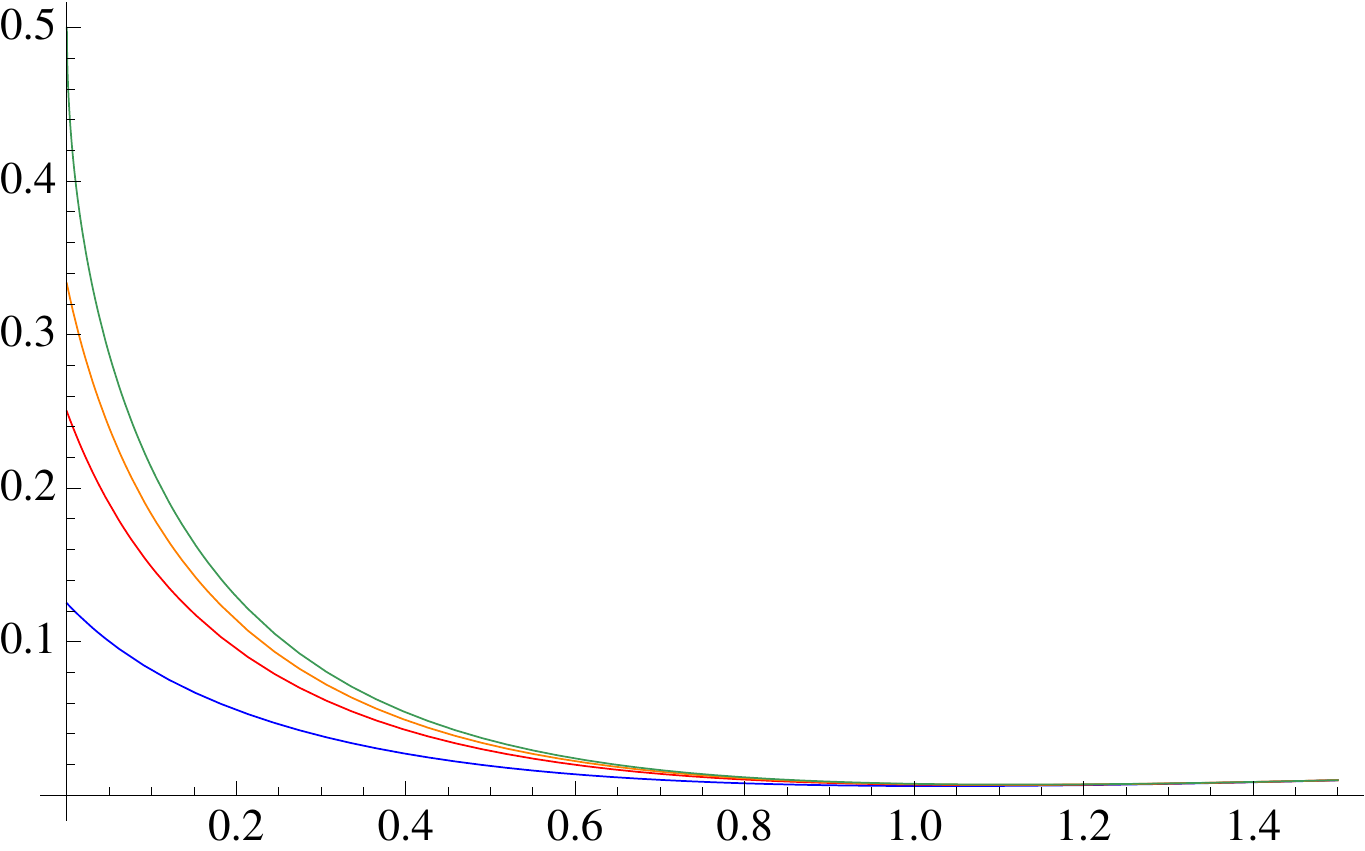}}
\put(85,45){\fbox{\includegraphics[scale=0.4]{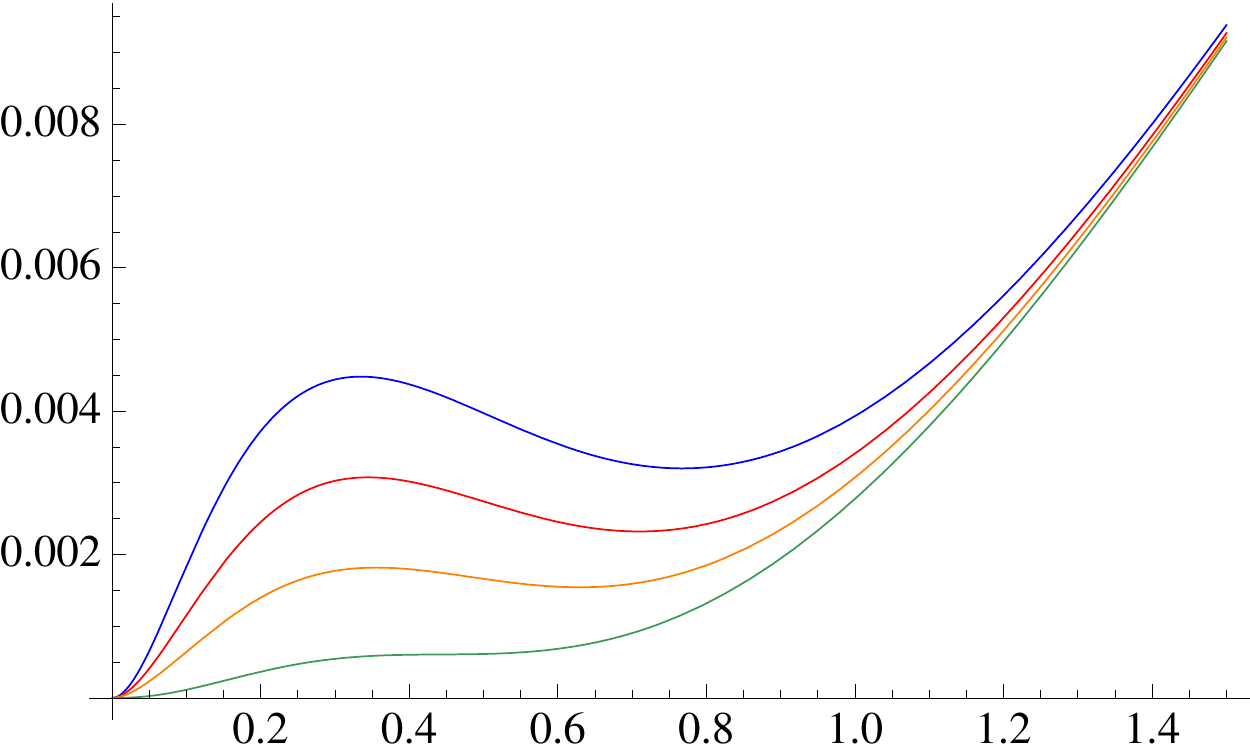}}}
\put(240,10){\makebox(0,0){\small{$|\alpha|^2$}}}
\put(210,53){\makebox(0,0){\scriptsize{$|\alpha|^2$}}}
\put(156,28){\makebox(0,0){\small{\textcolor{Green}{$\eta_1=\frac{1}{2}$}, \textcolor{orange}{$\eta_1=\frac{1}{3}$}, \textcolor{red}{$\eta_1=\frac{1}{4}$}, \textcolor{blue}{$\eta_1=\frac{1}{8}$}}}}
\put(120,150){\makebox(0,0){\small{(a) $P_\mathrm{err}^\mathrm{min}$ for different prior distributions}}}
\put(150,120){\makebox(0,0){(b) $P_\mathrm{err}^\mathrm{min}-P_E$}}
}
\put(0,-5){
\put(0,0){\includegraphics[scale=0.62]{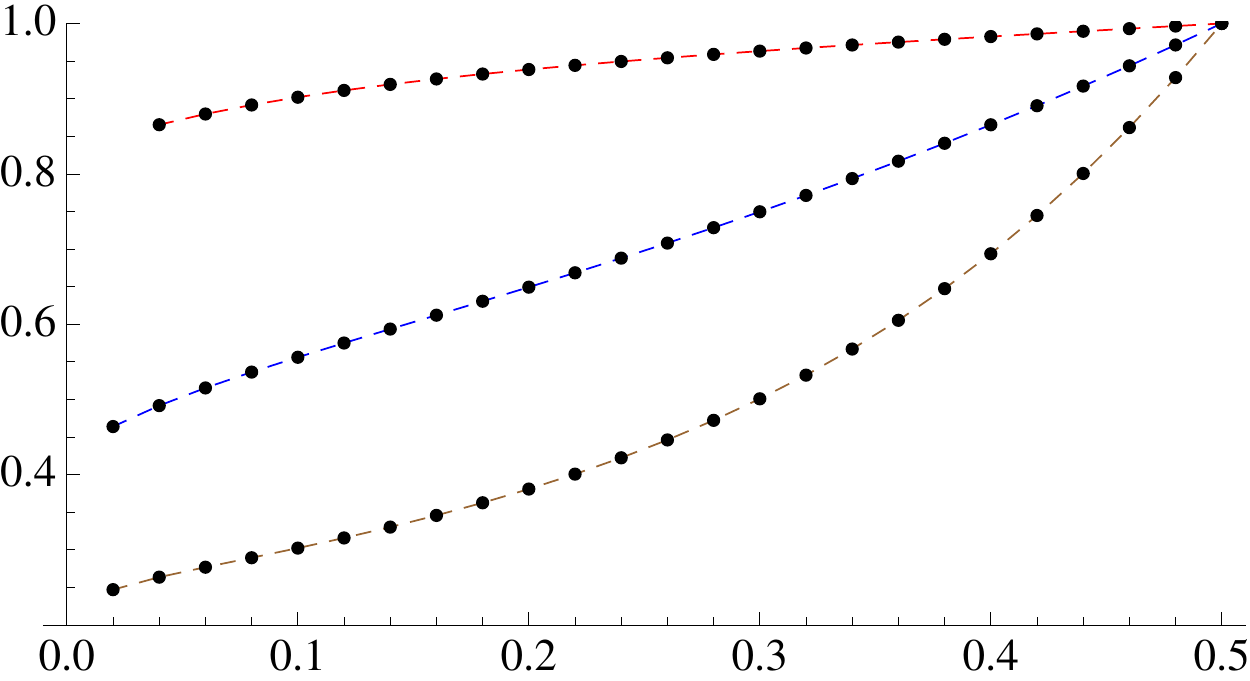}}
\put(235,8){\makebox(0,0){\small{$\eta_1$}}}
\put(110,126){\makebox(0,0){\small{(c) Plots of $\gamma_\mathrm{opt}$ against $\eta_1$}}}
\put(60,112){\makebox(0,0){$\alpha=1$}}
\put(85,78){\makebox(0,0){$\alpha=1/2$}}
\put(160,45){\makebox(0,0){$\alpha=1/4$}}
}
\end{picture}}
\caption{(a) Values of the error probability $P_\mathrm{err}^\mathrm{min}=\min_\Phi\frac{1}{2}[1-2D_\mathrm{tr}(\eta_1\rho_{A,\alpha},\eta_2\rho_{A,-\alpha})]$ for different prior distributions against the mean photon number $|\alpha|^2$. For the plots, the minimum values in the range $7.5<\Phi<9$ are used. The green, orange red and blue curves are for $\{\eta_1,\eta_2\}=\{\frac{1}{2},\frac{1}{2}\},\{\frac{1}{3},\frac{2}{3}\},\{\frac{1}{4},\frac{3}{4}\}$ and $\{\frac{1}{8},\frac{7}{8}\}$, respectively. (b) Difference between $P_\mathrm{err}^\mathrm{min}$ and the Helstrom bound for different prior probabilities; the color coding is the same as of (a).
(c) List plots of the optimal values of $\gamma$ against the prior probability $\eta_1$ for $\alpha=\{1,1/2,1/4\}$.}\label{fig:ErrorProbPrior}
\end{figure}
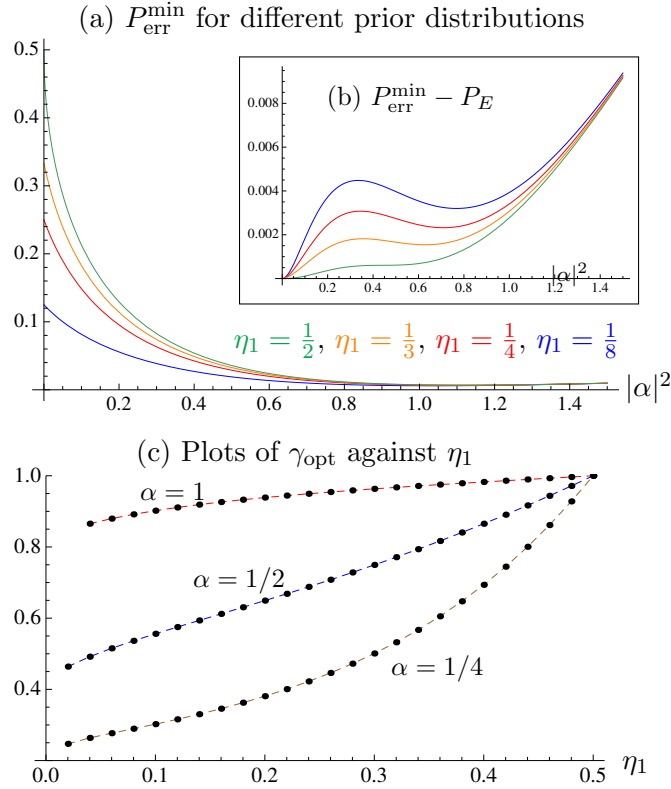

The which-way knowledge that Bob acquires from the outcomes of the projective measurement $\Pi(\gamma,\theta)$ is
\begin{eqnarray}
&&\big|\tr{\Pi(\gamma,\theta)(\eta_1\rho_{A,\alpha}-\eta_2\rho_{A,-\alpha})}\big|\\
&=&\frac{1}{1+\gamma^2}\sum_{n=0}^\infty \alpha_n^2\Big\{|\eta_1-\eta_2|\left[\cos^2(\Phi\sqrt{n})+\gamma^2\sin^2(\Phi\sqrt{n})\right]\nonumber\\[-2ex]
&&\hspace{14ex}+\frac{2\gamma\mathrm{Im}(\alpha e^{i\theta})}{\sqrt{n+1}}\cos(\Phi\sqrt{n})\sin(\Phi\sqrt{n+1})\Big\}.\nonumber
\end{eqnarray}
The optimal value of $\theta$ is the same as before because $\max[\mathrm{Im}(\alpha e^{i\theta})]=|\alpha|$ for $\theta=\pm\pi/2$, and it does not depend on any of the other parameters.
In the previous case when $\eta_1=\eta_2$, the maximum is at $\gamma=1$ and it is independent of the field amplitude $\alpha$. 
However, when $\eta_1\neq\eta_2$, the results indicate that the optimal value of $\gamma$ (denoted by $\gamma_\mathrm{opt}$) does depend on the field amplitude $\alpha$; see Fig.~\ref{fig:ErrorProbPrior}c. 
For $\eta_1<1/2$, the value of $\gamma_\mathrm{opt}$ increases monotonically as a function of $\eta_1$ for any fixed value of $\alpha$; for $\eta_1=1/2$, $\gamma_\mathrm{opt}$ converges to 1 for all $\alpha$; and, the function is symmetric about $\eta_1=1/2$ for $\eta_1>1/2$.
Figure~\ref{fig:ErrorProbPrior}c also indicates that the dependence of the optimal measurement on $\eta_1$ becomes weaker as $\alpha$ increases.

The minimum error probability $P_\mathrm{err}^\mathrm{min}$ is attained when $D_\mathrm{tr}(\eta_1\rho_{A,\alpha},\eta_2\rho_{A,-\alpha})$ is maximized over $\Phi$. Similarly to the case of equal priors, we find that for weak signal, the global maximum of the oscillatory function $D_\mathrm{tr}(\eta_1\rho_{A,\alpha},\eta_2\rho_{A,-\alpha})$ is given by its maximum around $\Phi=8$. For a different set of prior distributions, the minimum error probabilities $P_\mathrm{err}^\mathrm{min}$ as a function of the field strength $|\alpha|^2$ are shown in Fig.~\ref{fig:ErrorProbPrior}a, and their deviation from the Helstrom bound is shown in Fig.~\ref{fig:ErrorProbPrior}b. Our results demonstrate that, not only for states with equal priors but also for states with any arbitrary priors, the error probability attained using this nondestructive implementation is extremely close to the Helstrom bound. However, the deviation from the Helstrom bound is larger for signals with more biased prior distributions.

\subsection{The scheme for subsequent measurements}

In this subsection, we discuss the subsequent measurements for the case of equal priors, i.e., $\eta_1=\eta_2$, and do not discuss explicitly the case of unequal priors because it would follow the same arguments with only minor adaptions. Since the optimal measurement operators are projectors onto atomic states $\ket{\pm}$ for $\eta_1=\eta_2$, to obtain the post-measurement light states, we can rewrite the atom-light state $\ket{\Psi(t)}$ given by Eq.~(\ref{eq:Psit1}) in the basis of $\{\ket{+,n},\ket{-,n}\}$ instead of in its original basis of $\{\ket{\mathrm{g},n},\ket{\mathrm{e},n}\}$.
Corresponding to Eq.~(\ref{unitary}), for states $\{\ket{\psi_1},\ket{\psi_2}\}=\{\ket{\alpha},\ket{\!-\!\alpha}\}$, if the detector $\Pi_-=\ket{-}\bra{-}$ clicks the post-measurement light states are
\begin{eqnarray}
\sqrt{p_1}\ket{\varphi_1}&=&\frac{1}{\sqrt{2}}\sum_{n=0}^\infty\left[\cos(\Phi\sqrt{n})\alpha_n+\sin(\Phi\sqrt{n+1})\alpha_{n+1}\right]\ket{n},\\
\sqrt{r_2}\ket{\varphi_2}&=&\frac{1}{\sqrt{2}}\sum_{n=0}^\infty(-1)^n\left[\cos(\Phi\sqrt{n})\alpha_n-\sin(\Phi\sqrt{n+1})\alpha_{n+1}\right]\ket{n};\label{eq:PMStates1}
\end{eqnarray}
and if the detector $\Pi_+=\ket{+}\bra{+}$ clicks, the states are
\begin{eqnarray}
\sqrt{r_1}\ket{\phi_1}&=&\frac{1}{\sqrt{2}}\sum_{n=0}^\infty\left[\cos(\Phi\sqrt{n})\alpha_n-\sin(\Phi\sqrt{n+1})\alpha_{n+1}\right]\ket{n},\\
\sqrt{p_2}\ket{\phi_2}&=&\frac{1}{\sqrt{2}}\sum_{n=0}^\infty(-1)^n\left[\cos(\Phi\sqrt{n})\alpha_n+\sin(\Phi\sqrt{n+1})\alpha_{n+1}\right]\ket{n}.\;\;\;\label{eq:PMStates2}
\end{eqnarray}
The probabilities $\{p_1,p_2,r_1,r_2\}$ are the normalization factors of the four states above. The conditions $p_1=p_2=p$ and $r_1=r_2=r$ are automatically fulfilled because the factors $(-1)^n$ in Eqs.~(\ref{eq:PMStates1}) and~(\ref{eq:PMStates2}) do not affect the normalization of the states.

When detector $\Pi_-$ clicks, which happens with probability $P_{\Pi_-}=\eta_1p_1+\eta_2r_2$, Bob would have a confidence probability $c_1=\eta_1p_1/P_{\Pi_-}$ that the post-measurement state is $\ket{\varphi_1}$ and a confidence probability $1-c_1$ that the post-measurement state is $\ket{\varphi_2}$. Similarly, when projector $\Pi_+$ clicks, which happens with probability $P_{\Pi_+}=\eta_1r_1+\eta_2p_2$, Bob's confidence probabilities corresponding to post-measurement states $\ket{\phi_1}$ and $\ket{\phi_2}$ are, then, $c_2=\eta_1r_1/P_{\Pi_+}$ and $1-c_2$. In order to extract more information from the post-measurement states, Bob should design his subsequent measurements with these confidence probabilities as the new prior probabilities of the post-measurement states corresponding to Alice's state $\ket{\psi_1}$ and $\ket{\psi_2}$. This is illustrated by (\ref{eq:sequential2}) below.
\begin{eqnarray}
\left\{\begin{array}{l}\eta_1=\frac{1}{2}, \ket{\psi_1}\\[4pt]\eta_2=\frac{1}{2}, \ket{\psi_2}\end{array}\right.\hspace{-5pt}\begin{array}{l l}
\begin{array}{l} \\[10pt] \nearrow \end{array}
& \hspace{-3pt}\Pi_-
\left\{\begin{array}{c l}c_1=\eta_1p_1/P_{\Pi_-}, & \ket{\varphi_1},\\[6pt]1-c_1=\eta_2r_2/P_{\Pi_-}, & \ket{\varphi_2},\end{array}\right.\\[12pt] 
 &\;\;\mathrm{or}\\[2pt]
\begin{array}{l}\searrow \\[20pt] \end{array}
& \hspace{-3pt}\Pi_+
\left\{\begin{array}{c l}c_2=\eta_1r_1/P_{\Pi_+}, & \ket{\phi_1},\\[6pt]1-c_2=\eta_2p_2/P_{\Pi_+}, & \ket{\phi_2}.\end{array}\right.\end{array}\;\;\;
\label{eq:sequential2}
\end{eqnarray}

Since the post-measurement states are not identical and the result of the first measurement is built into the confidence probabilities as the new priors, subsequent measurements can always reduce the error probability as long as the measurement is not the identity operator. One can either design the subsequent measurement to be a direct measurement on the post-measurement light state or introduce more ancilla systems to interact with the post-measurement light to implement nondestructive measurements. The nondestructive subsequent measurements can be optimized in the same way as the first measurement, thus the details are omitted here.

However, if the optimal first measurement for discriminating binary light signals $\{\ket{\alpha},\ket{\!-\!\alpha}\}$ can be experimentally implemented with high precision, subsequent measurements might not even be necessary as the Helstrom bound is already almost saturated by one measurement. In practice, this depends on the precision requirement of the state discrimination task at hand and the experimental imperfections.
Moreover, in Ref.~\cite{Han:2017c}, we investigate the present non-destructive sequential discrimination scheme for the maximization of the mutual information rather than the minimization of the probability of error. We find that, for the present implementation, the information that is not successfully extracted by a measurement can be fully retrieved via subsequent measurements on the post-measurement states. This demonstrates that the non-destructive implementation is not only promising for the minimum error strategy but also for the maximization of mutual information.

\section{Discussion and Outlook}

\begin{figure}[t]
\centerline{
\begin{picture}(230,130)(0,0)
\small
\put(-15,28){\includegraphics[scale=0.46]{figure1part2.pdf}}
\put(54,60){\makebox(0,0){$\ket{\alpha}$}}
\put(20,60){\makebox(0,0){$\ket{\!-\!\alpha}$}}
\put(20,0){
\put(94,5){\includegraphics[scale=0.5]{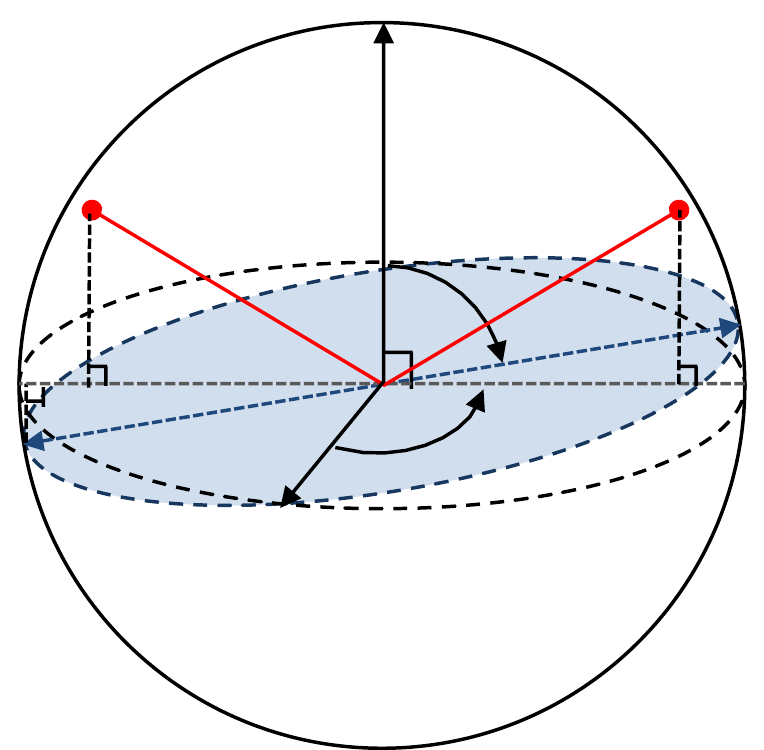}}
\put(208,54){\makebox(0,0){$\hat{\boldsymbol{y}}$}}
\put(218,66){\makebox(0,0){$\Pi_{\mathrm{opt},+}$}}
\put(83,50){\makebox(0,0){$\Pi_{\mathrm{opt},-}$}}
\put(154,116){\makebox(0,0){$\hat{\boldsymbol{z}}$, $\ket{\mathrm{g}}$}}
\put(150,-3){\makebox(0,0){$\ket{\mathrm{e}}$}}
\put(132,35){\makebox(0,0){$\hat{\boldsymbol{x}}$}}
\put(94,88){\makebox(0,0){$\rho_{A,\alpha}$}}
\put(210,85){\makebox(0,0){$\rho_{A,-\alpha}$}}
\put(174,72){\makebox(0,0){\scriptsize{$2\arctan\gamma_\mathrm{opt}$}}}
\put(158,46){\makebox(0,0){$\theta$}}
}
\end{picture}}
\caption{Scheme of the nondestructive implementation for the discrimination of binary coherent-state signals $\{\ket{\alpha},\ket{\!-\!\alpha}\}$ with unequal priors $\eta_1<\eta_2$. The atom is initially prepared in state $\ket{\mathrm{g}}$ and the optimal atomic measurement is the projective measurement $\Pi_\mathrm{opt}(\gamma_\mathrm{opt},\pm\pi/2)$ that gives $D_\mathrm{tr}(\eta_1\rho_{A,\alpha},\eta_2\rho_{A,-\alpha})=\big|\mathrm{tr}\{{\Pi_\mathrm{opt}(\eta_1\rho_{A,\alpha}-\eta_2\rho_{A,-\alpha})}\}\big|-\frac{1}{2}|\eta_1-\eta_2|$. One identifies the signals $\ket{\alpha}$ and $\ket{\!-\!\alpha}$ by the clicks of atomic measurements $\Pi_{\mathrm{opt},-}$ and $\Pi_{\mathrm{opt},+}$, respectively.}\label{fig6}
\end{figure}

The results of the last section demonstrate that the information encoded in binary coherent signals can be effectively transferred to a two-level ancilla atom via the Jaynes-Cummings interaction. 
If the atom is originally prepared in its ground state $\ket{\mathrm{g}}$, the Jaynes-Cummings interaction rotates its Bloch vector symmetrically about $\hat{\boldsymbol{x}}$ to the left by state $\ket{\alpha}$ and to the right by state $\ket{\!-\!\alpha}$, where the atomic state becomes mixed due to the atom-light entanglement. 
This symmetry in the atomic state rotation comes from the symmetry of both the Jaynes-Cummings Hamiltonian and the binary light signal. 
The optimal scheme for equal priors ($\eta_1=\eta_2$) corresponds to the maximum angle of rotation of the atomic state about $\hat{\boldsymbol{x}}$ and projective atomic measurement $\Pi_\pm$ that is along the $\hat{\boldsymbol{y}}$ direction; see Fig.~\ref{fig1}. For unequal priors, $\eta_1<\eta_2$ for example, the optimal measurement is rotated about $\hat{\boldsymbol{x}}$ such that the state with the larger prior probability, $\rho_{A,-\alpha}$ in this case, is detected with smaller error; see Fig~\ref{fig6}.

The key reason why the present scheme is superior to the schemes using standard photon detectors is twofold. 
First of all, our scheme fully explores the fact that the information is encoded in the phase of the weak coherent field amplitude rather than its quadrature or photon statistics. The atom-light interaction directly imprints the phase of the field amplitude onto the direction of the $\sigma_x$-rotation of the atomic state, such that the atomic measurement result reflects directly the quantum information encoded in the coherent signal. On the other hand, the standard schemes that discriminate the coherent states by measuring their field quadratures are trying to extract the quantum information of the states by measuring classical field properties, so they only work well to discriminate states with large amplitude (the `more classical' signals) but fail to work in the regime of weak coherent signals where the overlap of the signals is large.
Although the PNRD scheme measures the photon statistics which is a quantum feature of the states, it still does not directly yield the phases of the field amplitude.
Secondly, our scheme is nondestructive on the light field such that information not extracted in the first measurement can still be accessed with subsequent measurements. 
In fact, Ref.~\cite{Han:2017c} shows that no information is destroyed by the projective atomic measurement in this scheme. Any information that is not extracted by a measurement can be potentially attained by subsequent measurements.
This sequential measurement feature is lacking for all other schemes where photon detectors are used to directly measure the field because the measurements completely destroy the light states.

In Section 3, we chose the initial atomic state to be $\ket{\mathrm{g}}$. It is, in fact, the choice of our initial state not only because it is easy to prepare, but also because it is the optimal one among all possible initial states that could provide the maximum distinguishability of the atomic states after the atom-light interaction. This is due to the symmetry of the Jaynes-Cummings Hamiltonian. Thus, the minimum error probability $P_\mathrm{err}^\mathrm{min}$ attained with initial state $\ket{\mathrm{i}}=\ket{\mathrm{g}}$ yields a lower bound than the error probability attained with any other arbitrary initial state $\ket{\mathrm{i}}$. 
In addition, we did not discuss the feature that Bob could introduce a displacement operator $D(\beta)$ to the light state, as employed in many other schemes, before sending the light to interact with the atom. It can also be shown that such a displacement would not help in discriminating $\{\ket{\alpha},\ket{\!-\!\alpha}\}$ using our nondestructive scheme as it destroys part of the inherent symmetry of the problem.

Instead of the implementation of the optimal measurement for MESD demonstrated in this paper, the nondestructive strategy can also be adapted to the implementation of other kinds of state discrimination problems. For example, it can implement the unambiguous measurement that always discriminates one of the states perfectly and the other state with some error; the implementation of this scheme is discussed in the Appendix. We show that the rate of inconclusive results attained via the nondestructive implementation is reaching that of the perfect Kennedy receiver~\cite{Kennedy73}, and is much smaller than that of the experimentally implementable Kennedy receiver with imperfect PNRD detectors. However, to adapt our scheme to the implementation of the optimal unambiguous state discrimination that saturates the Ivanovic-Dieks-Peres (IDP) limit~\cite{IVANOVIC1987257, DIEKS1988303, PERES198819}, we need to introduce an ancilla system with at least three levels.

Other than the discrimination of binary coherent states, our scheme also offers alternative ways to discriminate other types of quantum signals using the nondestructive implementation. These alternative implementations can be particularly useful when the information is encoded in continuous variable states, where direct projective measurements on the systems are unavailable, such as the discrimination of squeezed states, and ternary or quaternary phase-shifted coherent states. For the example of the widely used quaternary phase-shifted coherent signals, Bob can entangle the field he receives to a four-dimensional ancilla system and choose four orthogonal projective measurements on the ancilla to establish an one-to-one correspondence between the measurement outcome and Alice's state;  or, alternatively, Bob can also use a lower-dimensional ancilla system and relate Alice's state to the measurement outcomes of a four-element POVM measurement. In general, there is much freedom in choosing the ancilla system and the measurements, and the optimal schemes need to be investigated to suit each particular discrimination problem at hand.

Last but not least, we would also like to point out that the nondestructive implementation scheme proposed in Ref.~\cite{Han:2017a} is not restricted to any type of physical ancilla system and unitary operation. In this work, we use a two-level atom (or an effective two-level atom) as the ancilla with the Jaynes-Cummings interaction and demonstrate the advantage of such a scheme over direct field measurements. In addition to the implementation based on ancilla atoms and atom-light interaction, other types of ancilla systems that can be entangled to the light signal effectively might also provide alternative ways to implement such nondestructive schemes. 
%\textcolor{red}{For example, in the ultra-strong coupling regime, the }
They are, however, not the subject of discussion in this paper.

\section{Summary}

In this paper, we have investigated the physical implementation of discriminating binary coherent signals through coupling the field to an ancilla atom via the Jaynes-Cummings interaction and projective measurements on the ancilla atom. In the present scheme, quantum information encoded in the phase of the coherent state is directly mapped onto the rotation of the atomic state which can be measured directly and easily. 
We have demonstrated that the error probability of this scheme can be extremely close to the Helstrom bound with optimized atom-light interaction and atomic measurement. The fact that the measurement is on the ancilla, hence nondestructive on the light signal, provides the possibility to perform a subsequent measurement on the post-measurement light states to further reduce the error probability. 
The proposed scheme for the implementation of near-optimal discrimination of weak coherent light states is not restricted by the use of any particular ancilla system as long as the interaction between the ancilla and the signal can be well described by the Jaynes-Cummings model. The experimental errors for each choice of the ancilla system need to be studied with respect to the actual experimental setups.

\appendix
\section{The nondestructive implementation of the Kennedy receiver}

In this Appendix, we study the implementation of the Kennedy receiver~\cite{Kennedy73, PhysRevLett.104.100505} using the nondestructive scheme described in this paper.

Prior to sending the light signal to interact with the ancilla atom, Bob can apply a displacement operator $D(\beta)$ to the coherent state Alice sent such that he would receive state $\ket{\psi_1}=\ket{\beta+\alpha}$ or $\ket{\psi_2}=\ket{\beta-\alpha}$. Then, the displaced coherent state is sent to interact with an ancilla atom prepared in its ground state $\ket{\mathrm{i}}=\ket{\mathrm{g}}$. After that, the atom is measured by projector $P_\mathrm{e}=\ket{\mathrm{e}}\bra{\mathrm{e}}$. If we set $\beta=\alpha$, the vacuum state $\ket{\psi_2}=\ket{0}$ does not affect the state of the atom at all, whereas the state $\ket{\psi_1}=\ket{2\alpha}$ can excite the atom from $\ket{\mathrm{g}}$ to $\ket{\mathrm{e}}$ with a maximum probability that roughly scales linearly in mean photon number $|\alpha|^2$ for small $|\alpha|$ and converges slowly to unity as $|\alpha|$ gets larger. 
This provides a physical implementation of the two-element POVM in the form of $\{\Pi_1\propto\ket{(-\alpha)^\perp}\bra{(-\alpha)^\perp},\Pi_2=1-\Pi_1\}$, where the ket $\ket{(-\alpha)^\perp}$ represents any state orthogonal to $\ket{\!-\!\alpha}$.
This is in the spirit of the so-called Kennedy receiver that unambiguously discriminates one of the signal states using a von Neumann measurement.

\begin{figure}[t]
\centerline{\setlength{\unitlength}{0.88pt}
\small
\begin{picture}(230,140)(0,0)
\put(0,00){\includegraphics[scale=0.52]{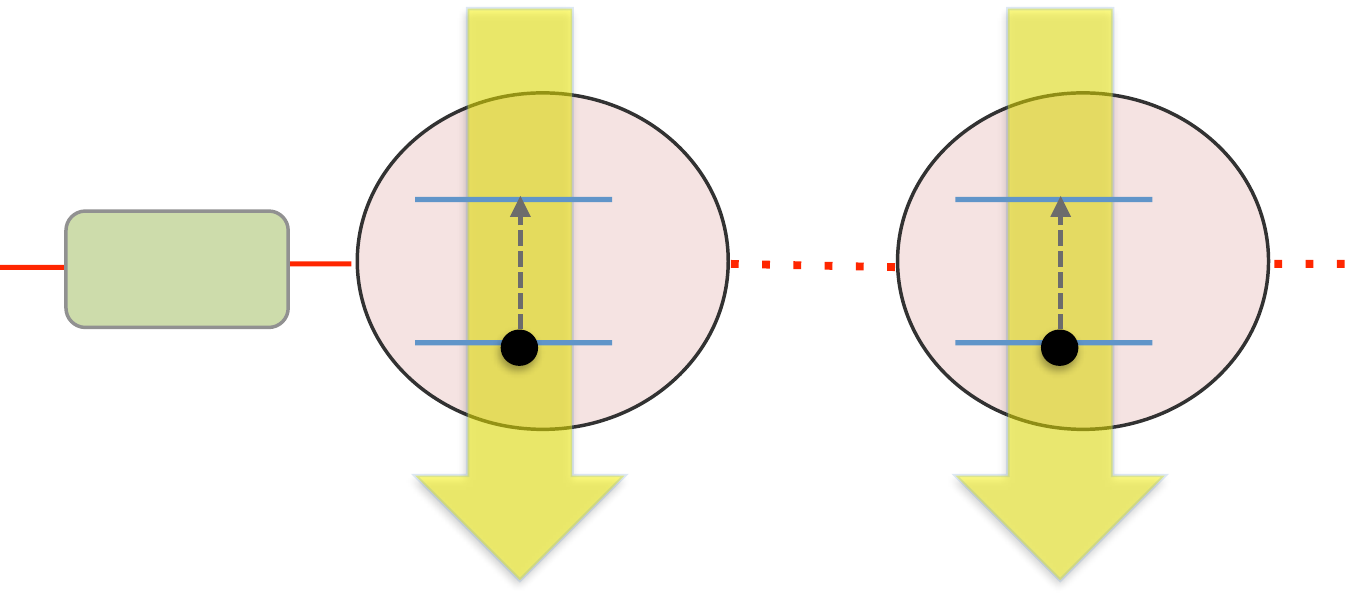}}
\put(30,56){\makebox(0,0){$D(\alpha)$}}
\put(88,108){\makebox(0,0){\scriptsize{ancilla atom}}}
\put(111,46){\makebox(0,0){$\ket{\mathrm{g}}$}}
\put(111,68){\makebox(0,0){$\ket{\mathrm{e}}$}}
\put(180,108){\makebox(0,0){\scriptsize{ancilla atom}}}
\put(204,46){\makebox(0,0){$\ket{\mathrm{g}}$}}
\put(204,68){\makebox(0,0){$\ket{\mathrm{e}}$}}
\put(135,120){\makebox(0,0){$\overbrace{\;\;\;\;\;\;\;\;\;\;\;\;\;\;\;\;\;\;\;\;\;\;\;\;\;\;\;\;\;}^{}$}}
\put(140,130){\makebox(0,0){$n$ subsequent measurements on the atoms}}
\put(138,20){\makebox(0,0){\begin{tabular}{c} \footnotesize{light is} \\ \footnotesize{entangled} \\ \footnotesize{with the atoms}\end{tabular}}}
\end{picture}}
\caption{Scheme of the nondestructive implementation of the Kennedy receiver that unambiguously discriminates one of the signal states. The light states $\{\ket{2\alpha},\ket{0}\}$, given by displacing states $\{\ket{\alpha},\ket{\!-\!\alpha}\}$ via displacement operator $D(\alpha)$, are sent to interact with a sequence of atoms in ground state $\ket{\mathrm{g}}$. The frequency and polarization of the coherent light state matches with that of the atomic transition, so that it can excite the atom from the ground state $\ket{\mathrm{g}}$ to the excited state $\ket{\mathrm{e}}$. Probing the atomic states completes the nondestructive measurement on the light states.}\label{fig2}
\end{figure}

In this setting, the rate of conclusive result for a single measurement is the probability that the atom is excited to state $\ket{\mathrm{e}}$ after the interaction with coherent state $\ket{2\alpha}$ times the prior probability of receiving state $\ket{\psi_1}$. 
If the measurement operator $P_\mathrm{e}$ has a click, corresponding to the detection of a fluorescence photon that can be done with a very high efficiency, Bob concludes with certainty that he received $\ket{\psi_1}$; whereas if no fluorescence photon is detected, the residue field can be sent to interact with another ground state ancilla atom. 
This sequential measurement feature is attributed to the nondestructive implementation of the POVM measurement, and this is the key difference between this scheme and the others using classical receivers. 
If a fluorescence photon is detected from the second ancilla atom, the conclusion that the state is $\ket{\psi_1}$ can be again made with certainty; otherwise the procedure of subsequent measurement continues. 
This simple strategy is illustrated in Fig.~\ref{fig2}. 
In order to minimize the number of subsequent measurements, optimization of the atom-light coupling to maximize the atomic excitation probability is essential.

\subsection{Maximizing the excitation probability}

If Bob receives coherent state $D(\alpha)\ket{\alpha}=\ket{2\alpha}$ and lets it interact with a two-level atom initially prepared in its ground state $\ket{\mathrm{g}}$, the ground and excited state populations of the atom at a later time can be attained from Eq.~(\ref{eq:Psit1}), i.e.,
\begin{equation}
|c_\mathrm{g}(\Phi)|^2=\sum_{n=0}^\infty|(2\alpha)_n|^2\cos^2\left(\Phi\sqrt{n}\right),\label{eq:population1g}\\
\end{equation}
\begin{equation}
|c_\mathrm{e}(\Phi)|^2=\sum_{n=0}^\infty|(2\alpha)_n|^2\sin^2\left(\Phi\sqrt{n}\right),\label{eq:population1e}
\end{equation}
where $|c_\mathrm{e}(\Phi)|^2=\mathrm{tr}_L\left\{|\expectn{\Psi(t)|e}|^2\right\}=1-|c_\mathrm{g}(\Phi)|^2$, the coefficients $(2\alpha)_n=e^{-2|\alpha|^2} \frac{(2\alpha)^n}{\sqrt{n!}}$, and the time dependence is implicitly hidden in the time integrated coupling strength $\Phi$ of Eq.~(\ref{eq:Phi}). 
On the other hand, if Bob receives state $D(\alpha)\ket{\!-\!\alpha}=\ket{0}$, the atom interacts with the vacuum and it stays in the ground state with $|c_\mathrm{g}(\Phi)|^2=1$. This indirect measurement scheme can be described as
\begin{equation}
\begin{array}{cll}
\ket{\alpha}&\rightarrow&U\ket{2\alpha}\ket{\mathrm{g}}=c_\mathrm{e}(\Phi)\ket{\varphi}\ket{\mathrm{e}}+c_\mathrm{g}(\Phi)\ket{\phi}\ket{\mathrm{g}},\\
\ket{\!-\!\alpha}&\rightarrow&U\ket{0}\ket{\mathrm{g}}=\ket{0}\ket{\mathrm{g}},
\end{array}
\label{unitary1}
\end{equation}
where $\ket{\varphi}$ and $\ket{\phi}$ are the corresponding post-measurement light states when $\ket{2\alpha}$ is sent. 

\begin{figure}[t]
\small
\centerline{\setlength{\unitlength}{1pt}
\begin{picture}(220,150)(0,0)
\put(0,-2){\includegraphics[scale=0.58]{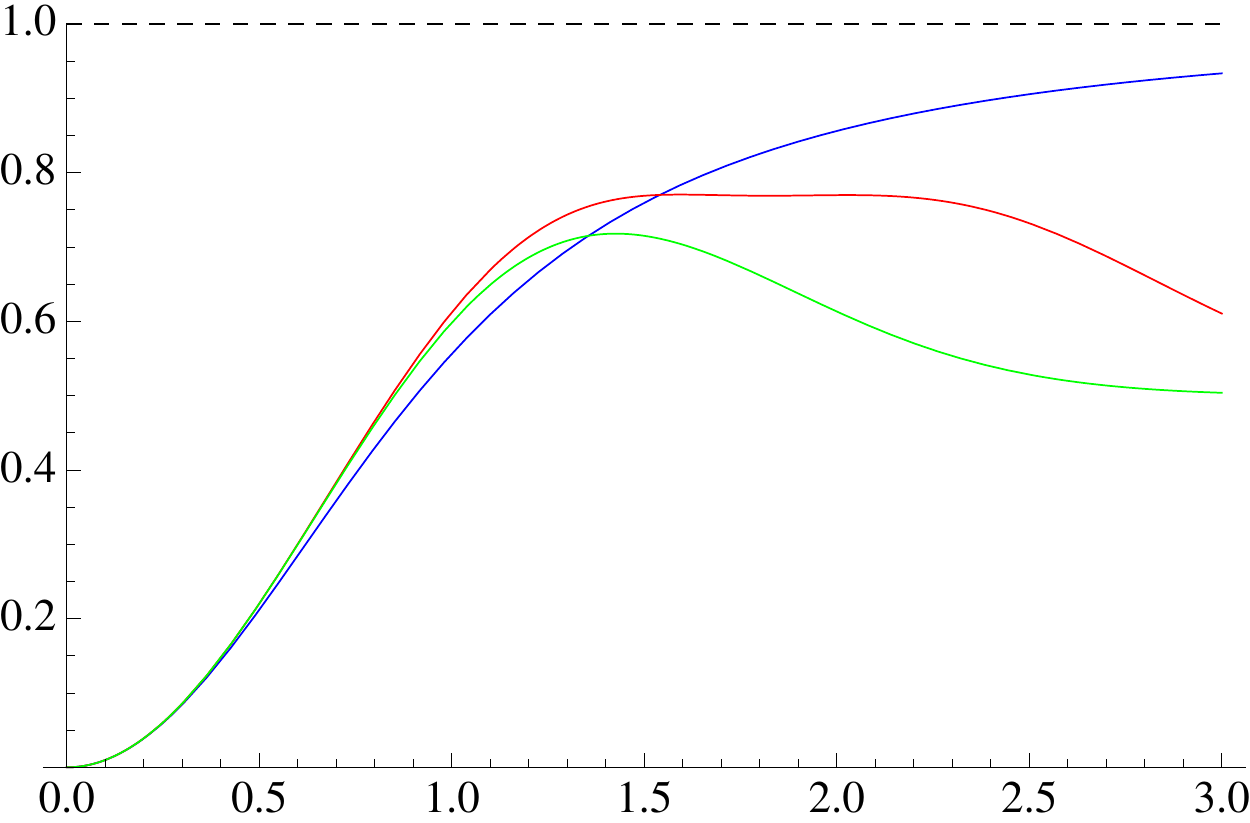}}
\put(70,112){\makebox(0,0){$\bar{p}_\mathrm{e}\equiv\max_\Phi[|c_\mathrm{e}|^2]$}}
\put(220,8){\makebox(0,0){$|2\alpha|$}}
\end{picture}}
\caption{Maximal values of the atomic excitation probability $\bar{p}_\mathrm{e}$ against field amplitude $|2\alpha|$ for the interaction between a ground state atom and the coherent light state $\ket{2\alpha}$.  The blue, green and red curves indicate the local maximal values obtained for $0<\Phi<2$, $7.8<\Phi<9$ and $30<\Phi<31$, respectively.}\label{figCe}
\end{figure}
Thus, optimal information on the light states can be extracted when the atomic excitation probability $|c_\mathrm{e}(\Phi)|^2$ in Eq.~(\ref{eq:population1e}) is maximized. We denote this maximum atomic excitation probability by
\begin{equation}
\bar{p}_\mathrm{e}\equiv\max_\Phi[|c_\mathrm{e}(\Phi)|^2]\,,
\end{equation}
where the maximization is over all positive values of $\Phi$.
In general, the excitation probability $|c_\mathrm{e}(\Phi)|^2$ is an oscillatory function in $\Phi$ with many local maxima and minima. 
For most values of $\alpha$ (roughly for $\alpha>0.8$), the global maximum of $|c_\mathrm{e}(\Phi)|^2$ is given by the local maximum attained for the smallest value of $\Phi$. 
For other values of $\alpha$, the first maximum is marginally smaller than the value of the global maximum attained around $\Phi=8$ and $\Phi=30$; see Fig.~\ref{figCe}. 
A large value of $\Phi(t)$ requires either a very long atom-light interaction duration or a very strong atom-light coupling that could be challenging to realize in practise.
Thus, in order to both increase experimental feasibility and reduce the complexity of the theoretical evaluation, we will only take the first maximum of $|c_\mathrm{e}(\Phi)|^2$ as the approximate value of $\bar{p}_\mathrm{e}$ in the following evaluations.

\subsection{The first measurement}

Bob's outcome is conclusive when the detector for $P_\mathrm{e}$ clicks and it happens with probability  $P_\mathrm{succ}=\eta_1 \bar{p}_\mathrm{e}$ (the overall success probability).
%If the atom is measured to be in state $\ket{\mathrm{g}}$, the original light state could either be $\ket{\alpha}$ presenting the post-measurement state as the vacuum $\ket{0}$ or $\ket{\!-\!\alpha}$ presenting the post-measurement state as $\ket{\phi}$. 
Thus, the failure probability,
\begin{equation}
Q\equiv 1-P_\mathrm{succ}=1-\eta_1 \bar{p}_\mathrm{e},\label{eq:inconclusiveP1}
\end{equation}
which is the rate of inconclusive outcome corresponding to no click for $P_\mathrm{e}$, is the probability that the atom remains in its ground state.
The Ivanovic-Dieks-Peres (IDP) limit~\cite{IVANOVIC1987257, DIEKS1988303, PERES198819} gives a lower bound of the failure probability, $Q^\mathrm{POVM}=2\sqrt{\eta_1\eta_2}e^{-2|\alpha|^2}$, that can be saturated using an optimized three-element POVM. However, since our scheme is only able to unambiguously discriminate one of the signal states, it is bounded by the optimal von Neumann measurement given by the perfect Kennedy receiver, $Q^\mathrm{Kennedy}=\eta_1e^{-4|\alpha|^2}+\eta_2\leq Q$, instead of being bounded by the IDP limit; see Fig.~\ref{fig:4USDMeasurement}.
Alternatively, if the figure of merit is the error probability, we have $P_\mathrm{err}^\mathrm{Kennedy}=\eta_1e^{-4|\alpha|^2}$ attained by associating the measurement outcome $\Pi_1$ with state $\ket{\alpha}$ and the measurement outcome $\Pi_2$ with state $\ket{\!-\!\alpha}$. For small $|\alpha|$ the error probability of the perfect Kennedy receiver is twice that of the Helstrom bound.

If the atom is found in state $\ket{\mathrm{e}}$, which is the conclusive outcome, it ends the discrimination procedure and no measurement on the post-measurement state $\ket{\varphi}$ is needed. 
However, if the atom is found in state $\ket{\mathrm{g}}$, the original light state could either be $\ket{\!-\!\alpha}$ or $\ket{\alpha}$ with corresponding post-measurement states $\ket{0}$ or $\ket{\phi}$. 
subsequent measurements to discriminate these two post-measurement states can reduce the rate of inconclusive result as long as these two states are not identical. 
By tracing out the atomic state after performing an operator $\ket{\mathrm{g}}\bra{\mathrm{g}}_A\otimes 1_L$ on the entangled state $\ket{\Psi(t)}$, the post-measurement state of light is
\begin{equation}
\ket{\phi}=\frac{1}{|c_\mathrm{g}(\Phi)|^2}\sum_{n=0}^\infty\cos(\Phi\sqrt{n})(2\alpha)_n\ket{n}\,,\label{eq:PostMState1}
\end{equation}
where the normalization factor $|c_\mathrm{g}(\Phi)|^2\geq1-\bar{p}_\mathrm{e}$ is given in Eq.~\eref{eq:population1g}.

\subsection{The subsequent measurements}

\begin{figure}[t]
\centerline{\setlength{\unitlength}{1pt}
\begin{picture}(250,170)(0,0)
\put(2,3){\includegraphics[width = 240pt]{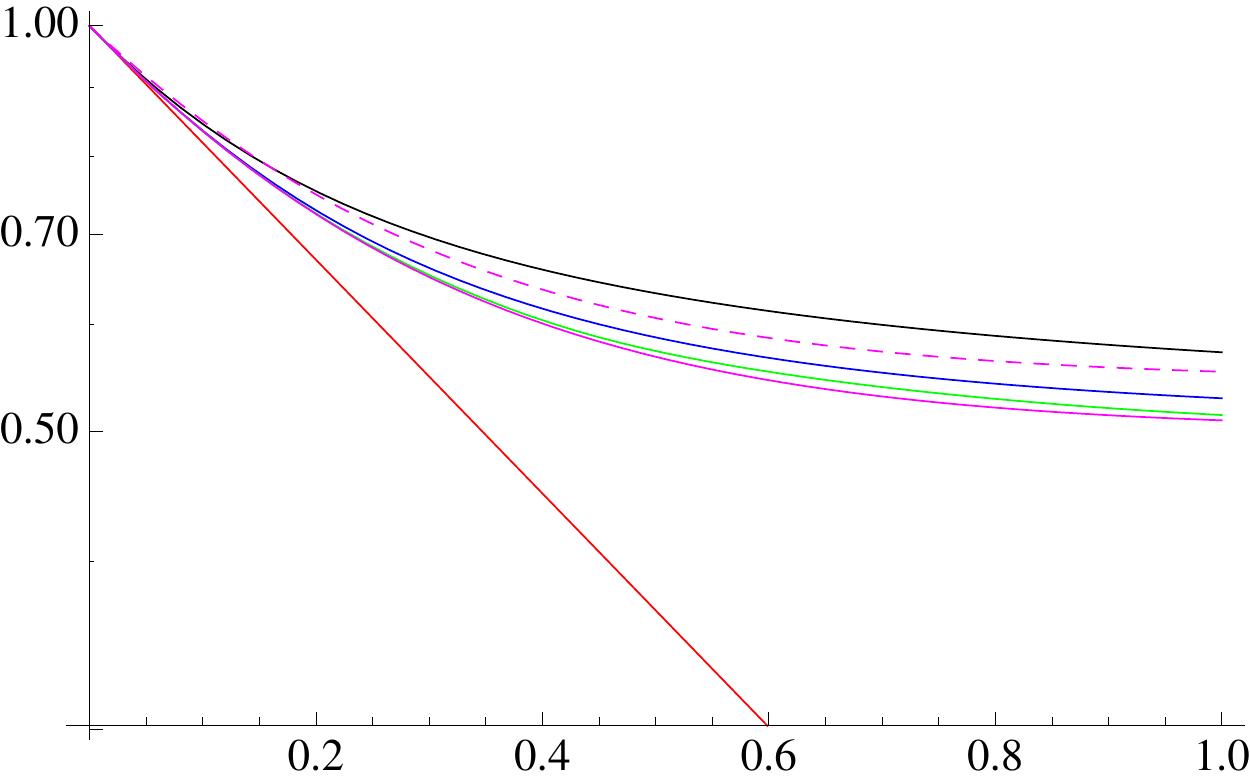}}
\put(15,158){\makebox(0,0){\small{$Q$}}}
\put(127,0){\makebox(0,0){\small{$|\alpha|^2$}}}
\put(100,40){\makebox(0,0){\small{$Q^\mathrm{POVM}$}}}
\put(170,68){\makebox(0,0){\small{$Q^\mathrm{Kennedy}$}}}
\end{picture}}
\caption{Failure probability $Q$ against $|\alpha|^2$ for unambiguous discrimination of coherent light states $\{\ket{\alpha},\ket{\!-\!\alpha}\}$ with equal priors $\eta_1=\eta_2=1/2$ after displacing the coherent state by $D(\alpha)$. The vertical axis is shown in logarithm scale. The black, blue and green curves show the minimum failure probabilities with the nondestructive implementation by detecting the excitation of ancilla atoms with one, two and three measurements, respectively. The lower bound of the failure probability given by the Ivanovic-Dieks-Peres limit is shown by the red curve, and the failure probabilities of the Kennedy receiver~\cite{Kennedy73} using PNRD with 100\% (solid) and 91\% (dashed) detector efficiencies are shown by the purple curves. Two subsequent measurements nearly saturate the ideal Kennedy limit.}\label{fig:4USDMeasurement}
\end{figure}

The second measurement aims at distinguishing the vacuum state $\ket{0}$ and state $\ket{\phi}$ given in Eq.~(\ref{eq:PostMState1}), when the first ancilla qubit is found in its ground state. To make the second measurement unambiguous, we again prepare a second ancilla atom in state $\ket{\mathrm{g}}$ and maximize the excitation probability when the field is in state $\ket{\phi}$. Let us denote the phase accumulated by the Jaynes-Cummings interaction with the second ancilla by
\begin{equation}
\Phi'(t')=\int_0^{t'}dt\,g'(t)\,,
\end{equation}
where $g'(t)$ is the coupling strength to the second ancilla at time $t$. The state of the atom-light system for the second measurement is
\begin{eqnarray}
\ket{\Psi(t)}=\frac{1}{|c_\mathrm{g}(\Phi)|^2}\sum_{n=0}^\infty&&\left[\cos\left(\Phi'\sqrt{n}\right)\cos(\Phi\sqrt{n})(2\alpha)_n \ket{\mathrm{g},n}\right.\nonumber\\
&&\left.-i\sin\left(\Phi'\sqrt{n}\right)\cos(\Phi\sqrt{n})(2\alpha)_{n}\ket{\mathrm{e},n-1}\right].
\end{eqnarray}\\
For a given $\Phi$ that optimizes the interaction with the first ancilla, we maximize the excitation probability of the second ancilla atom,
\begin{equation}
\left|c_\mathrm{e}'\left(\Phi,\Phi'\right)\right|^2=\sum_{n=0}^\infty \sin^2\left(\Phi'\sqrt{n}\right)\frac{\cos^2(\Phi\sqrt{n})|(2\alpha)_{n}|^2}{|c_{\mathrm{g},2\alpha}(\Phi)|^2}\,,
\end{equation}
and denote the maximum excitation probability for the second atom by 
$\bar{p}_\mathrm{e}'\equiv\,\mathrm{Max_{\Phi'}\big[|c_\mathrm{e}'(\Phi,\Phi')|^2\big]}$.
In this case, the second nondestructive measurement is described by the operation
\begin{equation}
\begin{array}{l}
U\ket{\phi}\ket{\mathrm{g}}=c_\mathrm{e}'\big(\Phi,\Phi'\big)\ket{\varphi'}\ket{\mathrm{e}}+c_\mathrm{g}'\big(\Phi,\Phi'\big)\ket{\phi'}\ket{\mathrm{g}},\\
U\ket{0}\ket{\mathrm{g}}=\ket{0}\ket{\mathrm{g}}.
\end{array}
\label{unitary2}
\end{equation}
The second measurement can unambiguously detect state $\ket{\phi}$ with probability $\bar{p}_\mathrm{e}'$ in the optimal case. Therefore, the total probability of failure given by the first two measurements is the probability that neither of the two ancilla atoms is found in the excited state when $\ket{\alpha}$ is sent, i.e.,
\begin{equation}
Q=1-\eta_1\left[\bar{p}_\mathrm{e}+(1-\bar{p}_\mathrm{e})\bar{p}_\mathrm{e}'\right].
\end{equation}

A subsequent measurement to discriminate the two residual states after the second measurement, $\ket{0}$ and $\ket{\phi'}$, can even further reduce the failure probability. Figure \ref{fig:4USDMeasurement} shows the minimum failure probability $Q$ as a function of the mean photon number $|\alpha|^2$ for such nondestructive implementations of the Kennedy receiver with up to three subsequent measurements. 
This figure clearly shows that the failure probability of our scheme with just two measurements already well surpasses that of the Kennedy receiver implemented experimentally using PNRDs with 91\% detector efficiency. In fact, our scheme with one measurement is comparable to the PNRD scheme with about 85\% detector efficiency.
Furthermore, Fig.~\ref{fig:4USDMeasurement} also indicates that the failure probability is reduced by every additional subsequent measurement. $Q$ for two measurements is significantly smaller than it is for only one measurement, but the reduction yielded by the third measurement is marginal. The failure rate for three of such subsequent nondestructive measurements almost saturates the failure rate for the Kennedy receiver with perfect detectors. 
However, the gap to the IDP limit is still large. This gap can potentially be reduced by using a three-level ancilla (or an ancilla with higher dimension) that is entangled to the light signal to implement an effective three-element POVM on the field instead of using a two-level ancilla.

\end{document}